\documentclass{article}

\usepackage{graphicx}
\usepackage{amssymb}
\def\abstract#1{\vskip 7mm 
        \begin{center}{\large Abstract}\par \smallskip
                \begin{minipage}[c]{12cm}
                        \small #1
                \end{minipage}
        \end{center}
}
\def\title#1{\begin{center}{\Large\bf #1}\end{center}}
\def\author#1{\vskip 5mm \begin{center}{#1}\end{center}}
\def\address#1{\begin{center}{\it #1}\end{center}}
\def\lsim{\mathrel{\rlap{\lower4pt\hbox{\hskip1pt$\sim$}}
    \raise1pt\hbox{$<$}}}                
\def\gsim{\mathrel{\rlap{\lower4pt\hbox{\hskip1pt$\sim$}}
    \raise1pt\hbox{$>$}}}                

\makeatletter
\@ifundefined{lesssim}{}{}
\@ifundefined{gtrsim}{}{}
\def\vereq#1#2{\lower3pt\vbox{\baselineskip1.5pt \lineskip1.5pt
\ialign{$\m@th#1\hfill##\hfil$\crcr#2\crcr\sim\crcr}}}
\makeatother

\begin{document}

\title{%
Statistical isotropy of CMB anisotropy from WMAP
}
\author{%
{\bf Tarun Souradeep}~\footnote{E-mail:tarun@iucaa.ernet.in} and
Amir Hajian~\footnote{E-mail:amir@iucaa.ernet.in}
}
\address{%
Inter-University Centre for   Astronomy and Astrophysics,\\ 
Post Bag 4, Ganeshkhind, Pune 411007,  India
}

\abstract{ The statistical expectation values of the temperature
fluctuations of cosmic microwave background (CMB) are assumed to be
preserved under rotations of the sky. We investigate the statistical
isotropy of the CMB anisotropy maps recently measured by the Wilkinson
Microwave Anisotropy Probe (WMAP) using bipolar spherical harmonic
power spectrum proposed in Hajian \& Souradeep 2003. The Bipolar Power
Spectrum (BiPS) is estimated for the full sky CMB anisotropy maps of
the first year WMAP data. The method allows us to isolate regions in
multipole space and study each region independently. This search shows
no evidence for violation of statistical isotropy in the
first-year WMAP data on angular scales larger than that corresponding
to $l\approx 60$ .}

\section{Introduction}

In standard cosmology, CMB anisotropy signal is expected to be
statistically isotropic, i.e., statistical expectation values of the
temperature fluctuations $\Delta T(\hat q)$ are preserved under
rotations of the sky. In particular, the angular correlation function
$C(\hat{q},\, \hat{q}^\prime)\equiv\langle\Delta T(\hat q)\Delta
T(\hat q^\prime)\rangle$ is rotationally invariant for Gaussian
fields. In spherical harmonic space, where $\Delta T(\hat q)=
\sum_{lm}a_{lm} Y_{lm}(\hat q)$ the condition of {\em statistical
isotropy} (SI) translates to a diagonal $\langle a_{lm} a^*_{l^\prime
m^\prime}\rangle=C_{l} \delta_{ll^\prime}\delta_{mm^\prime}$ where
$C_l$, the widely used angular power spectrum of CMB anisotropy. SI
CMB sky is essential for $C_l$ to be a complete description of
(Gaussian) CMB anisotropy and hence an adequate measure for comparing
with models. Hence, it is crucial to be able to determine from the
observed CMB sky whether it is a realization of a statistically
isotropic process, or not. The detection of statistical isotropy (SI)
violations in the CMB signal can have exciting and far-reaching
implication for cosmology.  For example, a generic consequence of
cosmic topology is the breaking of statistical isotropy in
characteristic patterns determined by the photon geodesic structure of
the manifold as probed by the CMB photons traveling to us from the
surface of last scattering over a distance comparable to the cosmic
horizon, $R_H$. On the other hand, SI violation could also arise from
foreground contamination, non-cosmological signals and be artifacts of
observational technique.

The first-year {\it Wilkinson Microwave Anisotropy Probe} ({\it WMAP})
observations are consistent with predictions of the concordance
$\Lambda$CDM model with scale-invariant and adiabatic fluctuations
which have been generated during the inflationary epoch
\cite{hin_wmap03, kogut_wmap03, sper_wmap03, page_wmap03,
peiris_wmap03}. After the first year of {\it WMAP} data, the SI of the
CMB anisotropy ({\it i.e.} rotational invariance of n-point
correlations) has attracted considerable attention.  Tantalizing
evidence of SI breakdown (albeit, in very different guises) has
mounted in the {\it WMAP} first year sky maps, using a variety of
different statistics. It was pointed out that the suppression of power
in the quadrupole and octopole are aligned \cite{maxwmap}.  Further
``multipole-vector'' directions associated with these multipoles (and
some other low multipoles as well) appear to be anomalously correlated
\cite{cop04, schw04}. There are indications of asymmetry in the power
spectrum at low multipoles in opposite hemispheres \cite{erik04a,
han04, nas04}. Possibly related, are the results of tests of
Gaussianity that show asymmetry in the amplitude of the measured genus
amplitude (at about $2$ to $3\sigma$ significance) between the north
and south galactic hemispheres \cite{par04, erik04b,
erik04c}. Analysis of the distribution of extrema in {\it WMAP} sky
maps has indicated non-gaussianity, and to some extent, violation of
SI \cite{lar_wan04}. However, what is missing is a common, well
defined, mathematical language to quantify SI (as distinct from non
Gaussianity) and the ability to ascribe statistical significance to
the anomalies unambiguously.

Since the observed CMB sky is a single realization of the underlying
correlation, the detection of SI violation or correlation patterns
pose a great observational challenge.  For statistically isotropic CMB
sky, the correlation function
\begin{equation}
C(\hat{n}_1,\hat{n}_2)\equiv C(\hat{n}_1\cdot\hat{n}_2) =
\frac{1}{8\pi^2}\int d{\mathcal R} C({\mathcal R}\hat{n}_1,\,
{\mathcal R}\hat{n}_2),
\label{avg_cth}
\end{equation}
where ${\mathcal R}\hat{n}$ denotes the direction obtained under the
action of a rotation ${\mathcal R}$ on $\hat{n}$, and $d{\mathcal R}$
is a volume element of the three-dimensional rotation group.  The
invariance of the underlying statistics under rotation allows the
estimation of $C(\hat{n}_1\cdot\hat{n}_2)$ using the average of the
temperature product $\widetilde{\Delta T}(\hat n) \widetilde{\Delta
T}(\hat n')$ between all pairs of pixels with the angular separation
$\theta$.  In the absence of statistical isotropy,
$C(\hat{n},\hat{n}')$ is estimated by a single product
$\widetilde{\Delta T}(\hat n)\widetilde{\Delta T}(\hat n')$ and hence
is poorly determined from a single realization. Although it is not
possible to estimate each element of the full correlation function
$C(\hat{n},\hat{n}')$, some measures of statistical anisotropy of the
CMB map can be estimated through suitably weighted angular averages of
$\widetilde{\Delta T}(\hat n)\widetilde{\Delta T}(\hat n')$. The
angular averaging procedure should be such that the measure involves
averaging over sufficient number of independent `measurements', but
should ensure that the averaging does not erase all the signature of
statistical anisotropy (as would happen in eq.~(\ref{avg_cth}) or
eq.~(\ref{bin_cth})).  Recently, we proposed the Bipolar Power
spectrum (BiPS) $\kappa_\ell$ ($\ell=1,2,3, \ldots$) of the CMB map as
a statistical tool of detecting and measuring departure from
SI~\cite{us_apjl,us_pascos} and reviewed in this article in
sec.~\ref{bips}. The non-zero value of the BiPS spectrum imply the
break down of statistical isotropy
\begin{equation} 
{\mathrm {\Huge STATISTICAL\,\,\,\, ISOTROPY}} \,\,\,\,\,\,\, \Longrightarrow \,\,\,\,\,\,\, 
\kappa_\ell\,=\,0 \,\,\,\,\,\,\, \forall \ell \ne 0.
\end{equation}
The BiPS is sensitive to structures and patterns in the underlying
total two-point correlation function \cite{us_apjl, us_pascos}.  The
BiPS is particularly sensitive to real space correlation patterns
(preferred directions, etc.) on characteristic angular scales. In
harmonic space, the BiPS at multipole $\ell$ sums power in
off-diagonal elements of the covariance matrix, $\langle a_{lm}
a_{l'm'}\rangle$, in the same way that the `angular momentum' addition
of states $l m$, $l' m'$ have non-zero overlap with a state with
angular momentum $|l-l'|<\ell<l+l'$. Signatures, like $a_{lm}$ and
$a_{l+n m}$ being correlated over a significant range $l$ are ideal
targets for BiPS. These are typical of SI violation due to cosmic
topology and the predicted BiPS in these models have a strong spectral
signature in the bipolar multipole $\ell$ space~\cite{us_prl}.  The
orientation independence of BiPS is an advantage since one can obtain
constraints on cosmic topology that do not depend on the unknown
specific orientation of the pattern ({\it{e.g.}}, preferred
directions).

The results of WMAP are a milestone in CMB anisotropy measurements
since it combines high angular resolution, high sensitivity, with
`full' sky coverage allowed by a space mission. The frequency coverage
allows for WMAP CMB sky maps to be foreground cleaned up to $l\sim
100$~\cite{maxwmap}. The CMB anisotropy map based on the WMAP data are
ideal for testing for statistical isotropy.

\section{Sources of Statistical Isotropy violation}

An observed map of CMB anisotropy, $\Delta T^{obs}_i$, contains the
true CMB temperature fluctuations, $\Delta T_i$, convolved with the
beam and buried into noise and foreground contaminations. The observed
map $\Delta T$ is related to the true map through this relation
\begin{equation}
 \Delta T^{obs}_i\,=\,\sum_j B_{ij} \Delta T_j\,+\,N_i,
\end{equation}
in which $B$ is a matrix that contains the information about the beam
smoothing effect and $\mathbf{n}$ is the contribution from
instrumental noise and foreground contamination. Hence, the observed
map is a realization of a Gaussian process with covariance
$C=C^T+C^N+C^{res}$ where $C^T$ is the theoretical covariance of the
CMB temperature fluctuations, $C^N$ is the noise covariance matrix and
$C^{res}$ is the covariance of residuals of foregrounds.  Breakdown of
statistical isotropy $C(\hat n, \hat n^\prime)\not\equiv C(\hat
n\cdot\hat n^\prime)$ can occur in any of these parts of the
correlation function. Broadly, these effects may be divided into two
kinds:
\begin{itemize}
\item{Theoretical signals:} These effects are theoretically motivated
and are intrinsic to the true CMB sky, ${\Delta T}$. We discuss two
examples of these effects, {\it i.e.} non-trivial cosmic topology and
primordial magnetic fields, in the next subsections.
\item{Observational artifacts:} In an ideally cleaned CMB map, the
true CMB temperature fluctuations are completely extracted from the
observed map. But this is not always true. Sometimes there are some
artifacts (related to $B$ or $N$) left in the cleaned map
which may in principle violate the SI. These effects are explained in
section \ref{artifacts}.
\end{itemize}

\subsection{Cosmic Topology \& Ultra-large scale structure}

\begin{figure}[h]
  \begin{center}
    \includegraphics[scale=0.6]{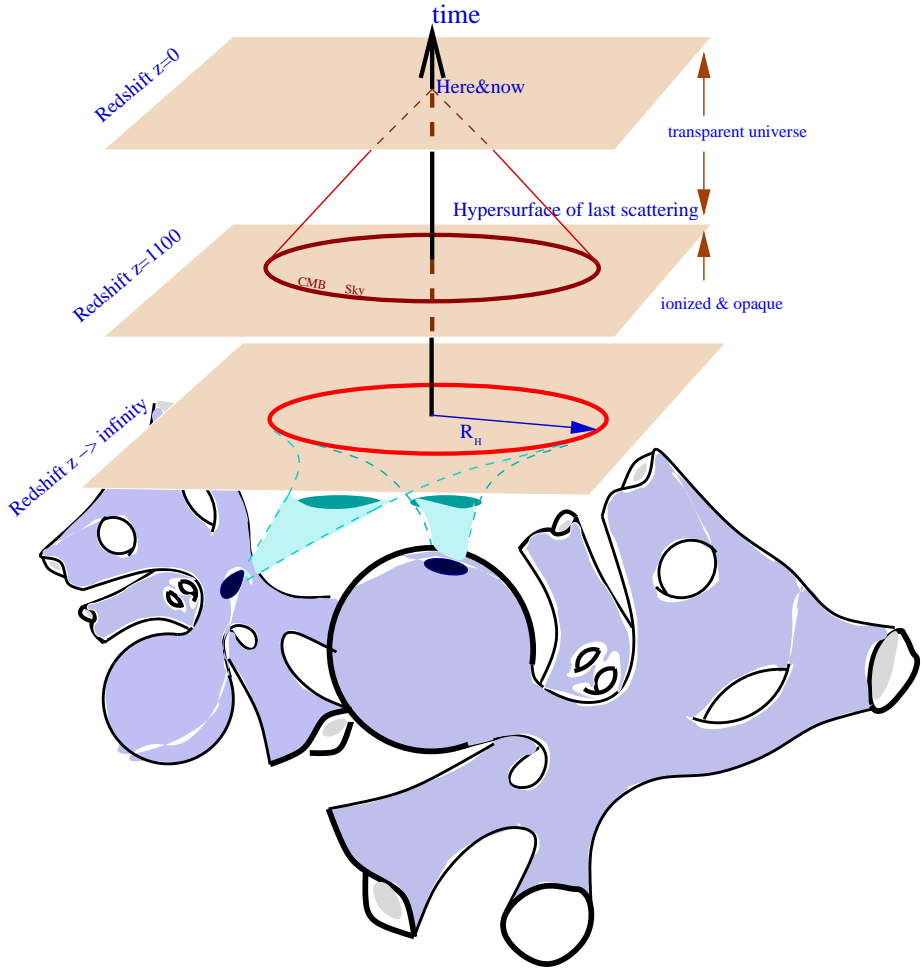}
  \end{center}
  \caption{A cartoon depicting a prevalent view within the
  inflationary paradigm. The observable universe corresponds to a
  small patch of a very complicated manifold that has been blown to
  cosmological scales during an inflationary epoch. Ultra-large scale
  structure could be observable if the the size of this patch is not
  much smaller that the scales of inhomogeneity and non-trivial
  topology.}
  \label{ulss_infl}
\end{figure}

The cosmic microwave background anisotropy is currently the most
promising observational probe of the global spatial structure of the
universe on length scales near to and even somewhat beyond the
`horizon' scale ($\sim c H_0^{-1}$).  Figure~\ref{ulss_infl} depicts a
prevalent view within the concept of inflation, that this relatively
smooth Hubble volume that we observe is perhaps a tiny patch of an
extremely inhomogeneous and complex spatial manifold.  The complexity
could involve non-trivial topology (multiple connectivity) on these
ultra-large scales. Within a general program to address the
observability of such a diverse global structure, a more well defined
and tractable path would be to restrict oneself to spaces of uniform
curvature (locally homogeneous and isotropic FRW models) but with
non-trivial topology; in particular, compact spaces which have
additional theoretical. 

 The question of size and the shape of our universe are very old
problems studied earlier~\cite{ell71,sok_shv74,got80,lac_lum95}. With
remarkable improvements in cosmological observations, in particular
the CMB anisotropy measurements, these questions have received
considerable attention over the past few
years~\cite{staro,stark98,levin98,
bps,angelwmap,coles-graca,cop04}. Although a multiply connected
universe sounds non-trivial, but there are theoretical motivations
\cite{linde, lev02} to favor a spatially compact universe. One
possibility to have a compact flat universe is the consideration of
multiply connected (topologically nontrivial) spaces.  The oldest way
of searching for global structure of the universe is by identifying
ghost images of local galaxies and clusters or quasars at higher
redshifts~\cite{lac_lum95}. This method can probe the topology of the
universe only on scales substantially smaller than the apparent radius
of the observable universe.  Another method to search for the shape of
the universe is through the effect on the cosmic density perturbation
fields.  For compact topologies, the two main effects on the CMB are:
(1) the breaking of statistical isotropy in characteristic patterns
determined by the photon geodesic structure of the manifold and (2) an
infrared cutoff in the power spectrum of perturbations imposed by the
finite spatial extent. More generally, in a universe with non-trivial
global spatial topology, the multiple connectivity of the space could
lead to observable characteristic angular correlation patterns in the
CMB anisotropy.

Over the past few years, many independent methods have been proposed
to search for evidence of a finite universe in CMB maps. These methods
can be classified in three main groups.
\begin{itemize}
\item Using the angular power spectrum of CMB anisotropies to probe
the topology of the Universe. The angular power spectrum, however, is
inadequate to characterize the peculiar form of the anisotropy
manifest in small universes of this type. Since nontrivial topology
breaks down SI, there is more information in a map of temperature
fluctuations than just the angular power spectrum \cite{levin98, bps,
us_prl}.
\item The second class of methods are direct methods that rely on
multiple imaging (or strong correlation features) of the CMB sky.
The most well known methods among these methods are S-map statistics
\cite{staro, angelwmap} and the search for circles-in-the-sky
\cite{circles}.

\item Third class of methods are indirect probes which deal with the
correlation patterns of the CMB anisotropy field by using an
appropriate combination of coefficients of the harmonic expansion of
the field \cite{coles-graca, donoghue, us_prl, cop04}. The Bipolar
power spectrum (BiPS) method is one of the strategies in this class.
\end{itemize}

\begin{figure}[h]
\includegraphics[scale=0.5,
angle=0]{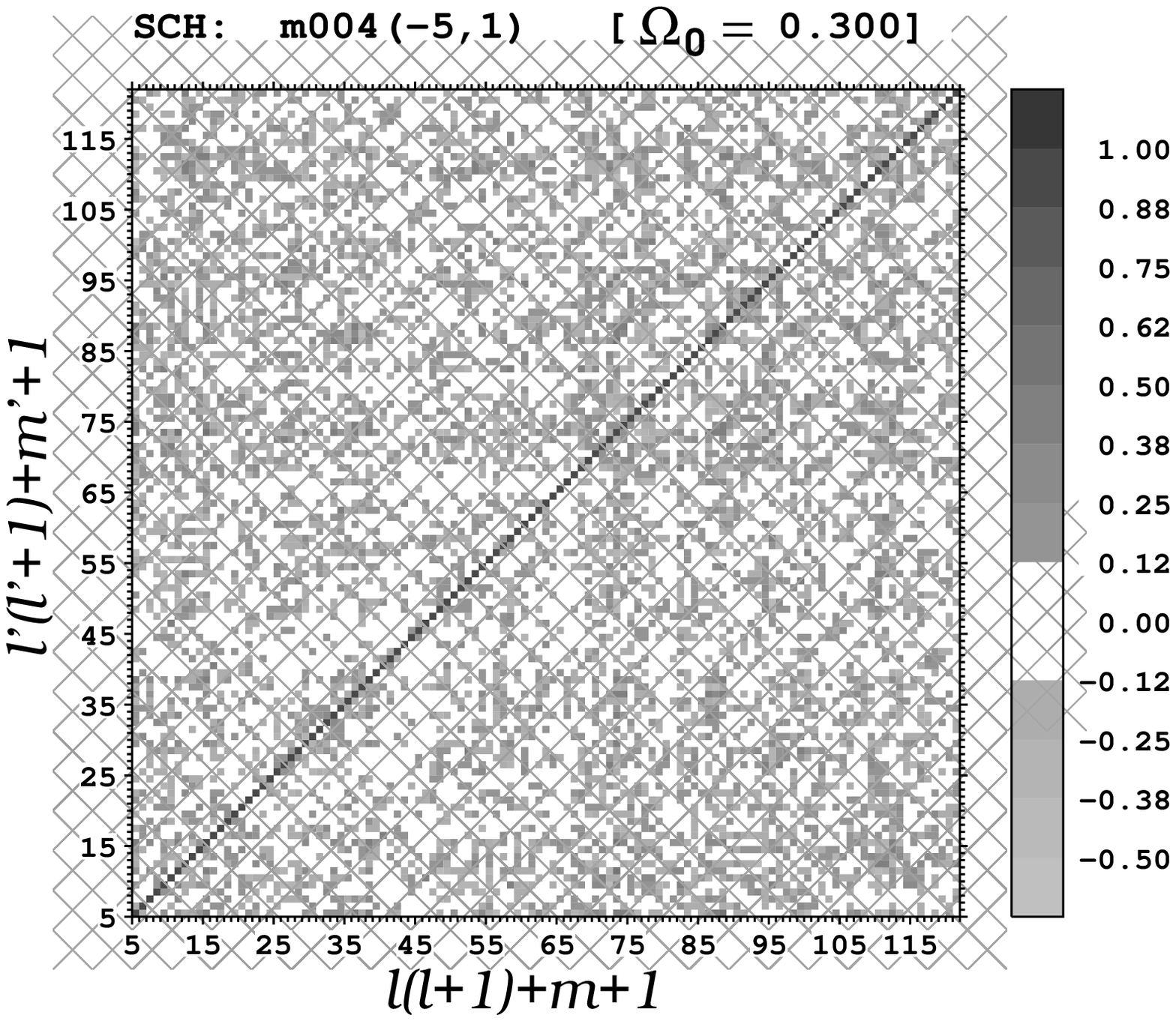}
\includegraphics[scale=0.5,
angle=0]{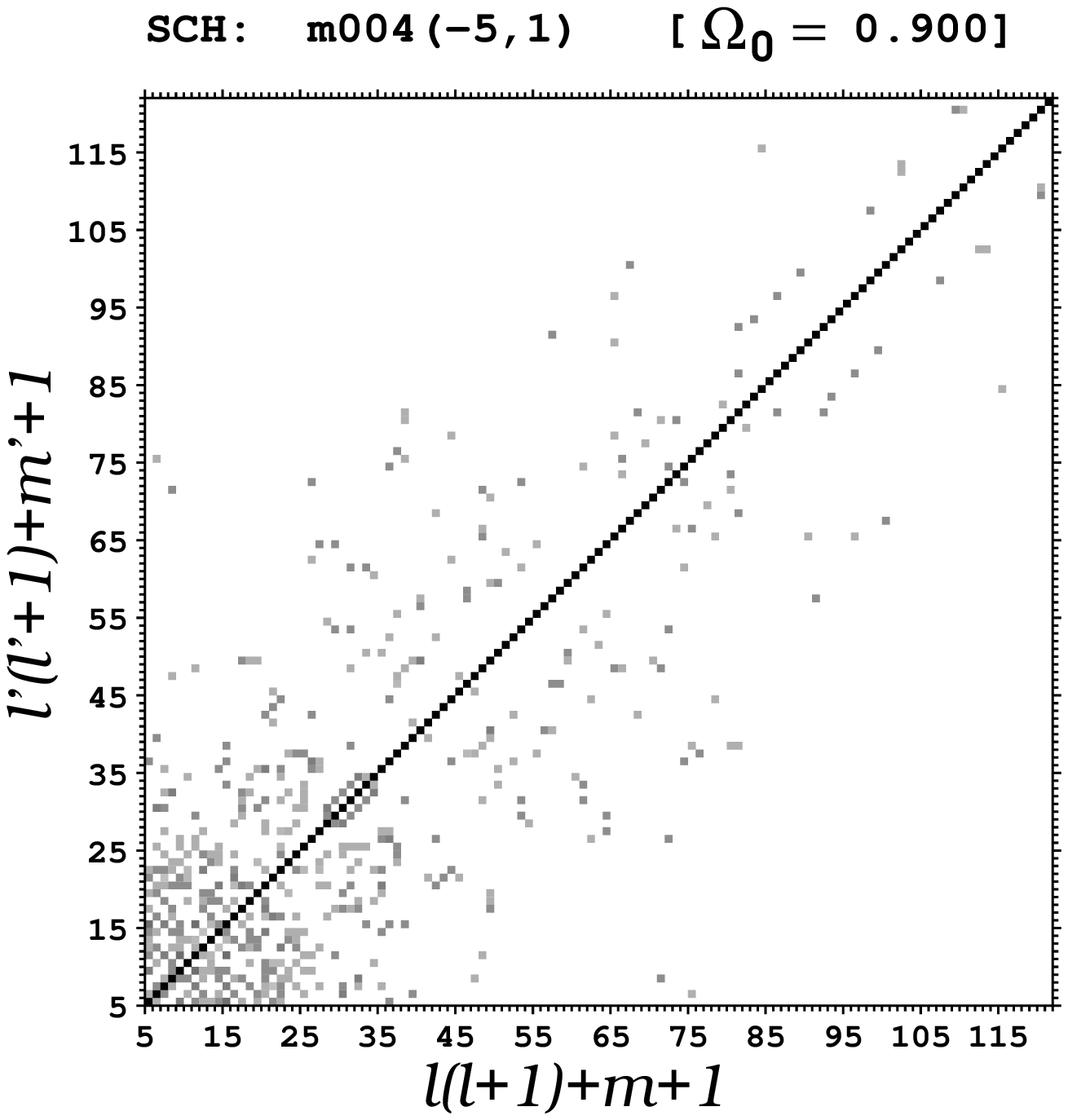}
\caption{The figure taken from~\protect\cite{bps} illustrates the
non-diagonal nature of the expectation values of $a_{\ell m}$ pair
products when the CMB anisotropy violates SI in two model compact
universe. The radical violation in the model on the left corresponds
to a small compact universe where CMB photons have traversed across
multiple times. The model on the left with mild violation of SI
corresponds to a universe of size comparable to the observable
horizon. For more details, see~\protect\cite{bps}}
\label{fig:cross_alm}
\end{figure}

The correlation patterns in CMB that lead to violation of SI implies
that imply $\langle\hat a_{l m} \hat a^*_{l m} \rangle$ has
off-diagonal elements. Figure~\ref{fig:cross_alm} taken from
~\cite{bps} shows the off-diagonal elements in the CMB correlation for
two compact universe models. BiPS gathers together the power in the
off-diagonal elements of $\langle\hat a_{l m} \hat a^*_{l m} \rangle$
as shown in Fig.~\ref{ALM}.

Using the fact that statistical isotropy is violated in compact spaces
one could use the bipolar power spectrum as a probe to detect the
topology of the universe. A simple example of is the BiPS signature of
a non-trivial topology can be given for a $T^3$ universe, where the
correlation function is given by
 \begin{equation}
C({\hat q,\hat q^\prime}) = L^{-3} \sum _{{\bf n}}
P_\Phi(k_{\bf n}) \,\,{\mathrm e}^{-i \pi 
(\epsilon_{\hat q} {\bf n}\cdot {\hat q} - \epsilon_{\hat q^\prime} {\bf n}\cdot 
{\hat q}^\prime)},
\label{C_tor}
\end{equation}
in which, ${\bf n}$ is 3-tuple of integers (in order to avoid
confusion, we use $\hat{q}$ to represent the direction instead of
$\hat{n}$), the small parameter $\epsilon_{\hat q} \le 1 $ is the
physical distance to the SLS along $\hat q$ in units of $L/2$ (more
generally, $\bar L/2$ where $\bar L= (L_1L_2L_3)^{1/3}$) and $L$ is
the size of the Dirichlet domain (DD). When $\epsilon$ is a small
constant, the leading order terms in the correlation function
eq.~(\ref{C_tor}) can be readily obtained in power series expansion in
powers of $\epsilon$.  For the lowest wave numbers $|{\mathbf n}|^2=1$
in a cuboid torus
\begin{eqnarray} 
C({\hat q,\hat q^\prime}) &\approx& 2 \sum_i P_\Phi({2\pi}/{L_i}) 
\cos(\pi\epsilon\beta_i\Delta q_i) \\ \nonumber
&\approx& C_0 \left[1 - \epsilon^2\, 
|{\mathbf \Delta q}|^2 + 3\,\epsilon^4 \, \sum_{i=1}^3
(\Delta q_i)^4   \right],
\label{appcorr}
\end{eqnarray} 
where $\Delta q_i$ are the components of ${\mathbf \Delta q} =\hat
q-\hat q^\prime$ along the three axes of the torus and $\beta_i = \bar
L /L_i$. From this, the non-zero $\kappa_\ell$ can be analytically
computed to be
\begin{eqnarray}
\frac{\kappa_0}{C_0^2}\, &=&\,\pi^2(1-4\epsilon^2 
+\frac{368}{15} \epsilon^4-\frac{288}{5}\epsilon^6+\frac{20736}{125}\epsilon^8)
\nonumber \\
\frac{\kappa_4}{C_0^2}\, &=&\, \frac{12288 \pi^2 }{875} \epsilon^8  
\end{eqnarray}
$\kappa_4$ has the information of the relative size of the Dirichlet
domain and one can use it to constrain the topology of the universe. A
detailed study of the BiPS signature of cosmic topology is given in
\cite{us_prl}.  These prediction allow us to constrain cosmic topology
using the BiPS measured in the observed CMB maps~\cite{us_dodeca}.

\subsection{Primordial Magnetic Fields}

Cosmological magnetic field, generated during an early epoch of
inflation~\cite{ratra1992, bamba2004}, can generate CMB
anisotropies~\cite{DKY}. The presence of a preferred direction due to
a homogeneous magnetic field background leads to non-zero off-diagonal
elements in the covariance matrix~\cite{gang}. This induces
correlations between $a_{l+1,m}$ and $a_{l-1,m}$ multipole
coefficients of the CMB temperature anisotropy field in the following
manner
\begin{equation}\label{alfven}
 \langle a_{lm}a^*_{l^\prime m^\prime}\rangle\,=\,
 \delta_{m,m'} [\delta_{l,l'}C_l+(\delta_{l+1,l'-1}+\delta_{l-1,l'+1}D_l)],
\end{equation}
where $D_l$ is the power spectrum of off-diagonal elements of the
covariance matrix. For a Harrison-Peebles-Yu-Zel'dovich
scale-invariant spectrum, $D_l$ behaves as $l^{-2}$. More precisely,
it is given by
 \begin{equation} 
 D_l\,=\,4\times 10^{-16} l^{-2} (\frac{B}{1nG})^4.
 \end{equation}
This clearly violates the statistical isotropy and gives rise to a
non-zero BiPS predictions for magnetic fields. This open the way to
use BiPS analysis on CMB maps to constrain or measure primordial
cosmic magnetic fields~\cite{us_magneticfield}.

\subsection{Observational Artifacts} \label{artifacts}

Foregrounds and observational artifacts (such as non-circular beam,
incomplete/non-uniform sky coverage and anisotropic noise) would also
manifest themselves as violations of SI.
\begin{itemize}
\item {\sf Anisotropic noise~:} The CMB temperature measured by an
  instrument is a linear sum of the cosmological signal as well as
  instrumental noise. The two point correlation function then has two
  parts, one part comes from the signal and the other one comes from
  the noise 

\begin{equation}C(\hat{n}_1,\, \hat{n}_2)\, =\, C^S(\hat{n}_1,\,
  \hat{n}_2)\,+\,C^N(\hat{n}_1,\, \hat{n}_2).  
\end{equation} 

Both signal and noise should be statistically isotropic to have a
statistically isotropic CMB map.  So even for a statistically
isotropic signal, if the noise fails to be statistically isotropic the
resultant map will turn out to be anisotropic. The noise matrix can
fail to be statistically isotropic due to non-uniform coverage. Also
if the noise is correlated between different pixels the noise matrix
could be statistically anisotropic. A simple example of this is the diagonal
 (but anisotropic) noise given by the following correlation
\begin{equation} 
C^N(\hat{n},\hat{n}')=\sigma^2(\hat{n}) \delta_{\hat{n}\hat{n}'}.
\end{equation}
This noise clearly violates the SI and will lead to a non-zero BiPS
given by
\begin{equation} 
\kappa_{\ell}=\sum_{m=-\ell}^{\ell}{|f_{\ell m}|^2},
\end{equation}
where $f_{\ell m}$ are spherical harmonic transform of the noise, $f_{\ell m}=\int{\mathrm{d} \Omega_{\hat{n}}Y_{\ell m}^{*}(\hat{n})\sigma^2(\hat{n})}$. 

\item {\sf The effect of non-circular beam~:} In practice when we deal
  with  data, it is necessary to take into account the {\it
    instrumental response}. The instrumental response is nothing but
  the beams width and the form of the beam and can be taken into
  account by defining a beam profile function $B(\hat{n},\,
  \hat{n}')$. Here $\hat{n}$ denotes the direction to the center of
  the beam and $\hat{n}' $ denotes the direction of the incoming
  photon. The temperature measured by the instrument is given by 

\begin{equation} 
  \Delta \tilde{T}(\hat{n})\,=\,\int \Delta T (\hat{n}')B(\hat{n},\,
  \hat{n}')d\Omega_{\hat{n}'}\,. \end{equation} Using this relation to
  calculate the correlation function $\tilde{C}(\hat{n}_1,\,
  \hat{n}_2)\, =\, \langle \Delta \tilde{T}(\hat{n}_1) \Delta
  \tilde{T}(\hat{n}_2)\rangle$ one would get \begin{eqnarray}
  \tilde{C}(\hat{n}_1,\, \hat{n}_2)\,&=&\,\int d\Omega_{\hat{n}'} \int
  d\Omega_{\hat{n}''} \langle \Delta T(\hat{n}') \Delta
  T(\hat{n}'')\rangle B(\hat{n}_1,\, \hat{n}')B(\hat{n}_2,\,
  \hat{n}'') \\ \nonumber &=&\,\int d\Omega_{\hat{n}'} \int
  d\Omega_{\hat{n}''} C(\hat{n}',\hat{n}'') B(\hat{n}_1,\,
  \hat{n}')B(\hat{n}_2,\, \hat{n}'').  \end{eqnarray} Only for a
  circular beam where $B(\hat{n},\, \hat{n}')\, \equiv
  \,B(\hat{n}\cdot \hat{n}')$, the correlation function is
  statistically isotropic, $\tilde{C}(\hat{n}_1,\,
  \hat{n}_2)\,\equiv\,\tilde{C}(\hat{n}_1 \cdot \hat{n}_2)$.
  Breakdown of SI is obvious since even $C_l$ get mixed for a
  non-circular beam, $\tilde{C}_l=\sum_{l'}{A_{ll'}C_{l'}}$
  \cite{beam}. Non-circularity of the beam in CMB anisotropy
  experiments is becoming increasingly important as experiments go for
  higher resolution measurements at higher sensitivity.
\item{\sf Mask effects~:} Many experiments map only a part of the sky.
  Even in the best case, contamination by galactic foreground
  residuals make parts of the sky unusable.  The incomplete sky or
  mask effect is another source of breakdown of SI.  But, this effect
  can be readily modeled out. The effect of a general mask on the
  temperature field is as follows
\begin{equation} 
\Delta T^{masked}(\hat{n})\, =\, \Delta T(\hat{n}) W(\hat{n}),
\end{equation}
where $W(\hat{n})$ is the mask function. One can cut different parts of the sky by choosing appropriate mask functions. Masked $a_{lm}$ coefficients can be computed from the masked temperature field,
\begin{eqnarray}  \label{maskedalm}
a_{lm}^{masked}
\,&=&\,\int{\Delta T^{masked}(\hat{n}) Y_{lm}^*(\hat{n}) d\Omega_{\hat{n}}}
\\ \nonumber
&=&\, \sum_{l_1 m_1}{a_{l_1 m_1} \int{Y_{l_1 m_1}(\hat{n}) Y_{lm}^*(\hat{n}) W(\hat{n}) d\Omega_{\hat{n}}}}.
\end{eqnarray} 
Where $a_{l_1 m_1}$ are spherical harmonic transforms of the original temperature field. We can expand $W(\hat{n})$ in spherical harmonics as well
\begin{equation}
W(\hat{n})\,=\,\sum_{l m}{w_{lm} Y_{l m}(\hat{n})},
\end{equation} 
and after substituting this into eq.~(\ref{maskedalm}) it is seen that  the
masked $a_{lm}$ is given by the effect of a kernel $K_{lm}^{l_1 m_1}$ on original $a_{lm}$ \cite{prunet}
\begin{equation} 
a_{lm}^{masked}\,=\, \sum_{l_1 m_1}{a_{l_1 m_1} K_{lm}^{l_1 m_1}}.
\end{equation}
The kernel contains the information of our mask function and is defined by
\begin{eqnarray} 
K_{lm}^{l_1 m_1}\,&=&\,\sum_{l_2m_2}{w_{l_2m_2}\int{Y_{l_1 m_1}(\hat{n}) Y_{l_2 m_2}(\hat{n})Y_{lm}^*(\hat{n}) d\Omega_{\hat{n}}}}
\\ \nonumber
&=&\,\sum_{l_2m_2}{w_{l_2m_2}\sqrt{\frac{(2l_1+1)(2l_2+1)}{4\pi(2l+1)}}C_{l_10l_20}^{l0}C_{l_1m_1l_2m_2}^{lm}}.
\end{eqnarray}
The covariance matrix of a masked sky will no longer have the diagonal
form because of the action of the kernel
\begin{eqnarray} 
\langle a^{masked}_{lm}a^{masked\, *}_{l^\prime m^\prime}\rangle
\,&=&\,\langle a_{l_1m_1}a^{*}_{l^\prime_1 m^\prime_1}\rangle 
K_{lm}^{l_1 m_1}K_{l^\prime m^\prime}^{l^\prime_1 m^\prime_1}\\ \nonumber
\,&=&\,C_{l_1}\delta_{l_1l^\prime_1}\delta_{m_1m^\prime_1}K_{lm}^{l_1 m_1}K_{l^\prime m^\prime}^{l^\prime_1 m^\prime_1}\\ \nonumber
\,&=&\,\sum_{l_1,m_1}{C_{l_1}K_{lm}^{l_1 m_1} K_{l^\prime m^\prime}^{l_1 m_1}}.
 \end{eqnarray} 
This clearly violates the SI and results a non-zero BiPS for masked CMB skies. 
 In the next section we apply a galactic mask to ILC map and  show that signature of this mask on BiPS is a rising tail at low $\ell, (\ell < 20)$. 
\item{\sf Residuals from foreground removal~:} Besides the
  cosmological signal and instrumental noise, a CMB map also contains
  foreground emission such as galactic emission, etc. The foreground
  is usually modeled out using spectral information. However,
  residuals from foreground subtractions in the CMB map will violate
  SI. Interestingly, BiPS does sense the difference between maps with
  grossly different emphasis on the galactic foreground. As shown in
  \cite{haj_sour05b} the BiPS of a Wiener filtered map shows a
  signal very similar to that of a galactic cut sky. This can be
  understood if one writes the effect of the Wiener filter as a weight
  on the `contaminated' galactic regions of the map.
\begin{equation}
\Delta T^{W}(\hat{n})\, =\, \Delta T(\hat{n}) (1+W(\hat{n})).
\end{equation}
This explains the similarity between a cut sky and a Wiener filtered map. 
The effect of foregrounds on BiPS still needs to be studied more carefully.
\end{itemize}

\section{The Bipolar Power Spectrum (BiPS)} 
\label{bips}

Two point correlation of the CMB anisotropy is given by ensemble
average, but there is only one observable CMB sky. Hence, the ensemble
average is meaningless unless the CMB sky is SI, when the two point
correlation function $C(\theta)$ can be well estimated as in
eq.~(\ref{bin_cth}) by the average product of temperature fluctuations
over all pairs of directions $\hat n_1$ and $\hat n_2$ whose angular
separation is $\theta$. In particular, for CMB temperature map
$\widetilde{\Delta T}(\hat n_i)$ defined on a discrete set of points
on celestial sphere (pixels) $\hat n_i$ ($i=1,\ldots,N_p$)
\begin{equation}
\tilde C(\theta)\,=\, \sum_{i,j=1}^{N_p} \widetilde{\Delta T}(\hat
n_i) \widetilde{\Delta T}(\hat n_j) \delta( \cos\theta - \hat n_i
\cdot \hat n_j)\,,
\label{bin_cth}
\end{equation}
is an estimator of the correlation function $C(\theta)$ of an
underlying SI statistics.
If the statistical isotropy is violated the estimate of the correlation
function from a sky map given by a single temperature product 
\begin{equation} 
\tilde C(\hat{n}_1, \hat{n}_2)\,=\, \Delta T(\hat{n}_1) \Delta
T(\hat{n}_2) 
\end{equation} 
is poorly determined.

Although it is not possible to estimate each element of the full
correlation function $C(\hat{n}_1,\hat{n}_2)$, some measures of
statistical isotropy of the CMB map can be estimated through suitably
weighted angular averages of $\Delta T(\hat{n}_1) \Delta
T(\hat{n}_2)$. The angular averaging procedure should be such that the
measure involves averaging over sufficient number of independent
measurements , but should ensure that the averaging does not erase all
the signature of statistical anisotropy. Another important desirable
property is that measure be independent of the overall orientation of
the sky. Based on these considerations, we have proposed a set of
measures of statistical isotropy~\cite{us_apjl} 
 \begin{equation}\label{kl}
 \kappa^{\ell}\,=\, (2l+1)^2 \int d\Omega_{n_1}\int d\Omega_{n_2} \,
 [\frac{1}{8\pi^2}\int d{\mathcal R} \chi^{\ell}({\mathcal R})\, C({\mathcal R}\hat{n}_1,\, {\mathcal R}\hat{n}_2)]^2.
 \end{equation}
 In the above expression, $ C({\mathcal R}\hat{n}_1,\, {\mathcal
 R}\hat{n}_2)$ is the two point correlation at ${\mathcal
 R}\hat{n}_1\,$ and $ {\mathcal R}\hat{n}_2$ which are the coordinates
 of the two pixels $\hat{n}_1\,$ and $\hat{n}_2$ after rotating the
 coordinate system through an angle $\omega$ where $(0\leq \omega \leq
 \pi)$ about the axis ${\bf n}(\Theta, \Phi)$.  The direction of this
 rotation axis ${\bf n}$ is defined by the polar angles $\Theta$ where
 $(0\leq \Theta \leq \pi)$ and $\Phi$, where $(0\leq \Phi \leq
 2\pi)$. $\chi^{\ell}$ is the trace of the finite rotation matrix in
 the $\ell M$-representation
 \begin{equation}\label{chil} 
 \chi^{\ell}({\mathcal R})\,=\,\sum_{M=-\ell}^{\ell} D_{MM}^{\ell}({\mathcal R}),
 \end{equation}
which is called the {\it characteristic function}, or the character of
the irreducible representation of rank $\ell$. It is invariant under
rotations of the coordinate systems. Explicit forms of
$\chi^{\ell}({\mathcal R})$ are simple when ${\mathcal R}$ is
specified by $\omega,\, \Theta,\,\Phi$, then $\chi^{\ell}({\mathcal
R})$ is completely determined by the rotation angle $\omega$ and it is
independent of the rotation axis ${\bf n}(\Theta, \Phi)$,
 \begin{eqnarray} 
 \chi^{\ell}({\mathcal R})\,&=&\,\chi^{\ell}(\omega)\,\\ \nonumber
 &=&\,\frac{\sin{[(2\ell+1)\omega/2]}}{\sin{[\omega/2]}}\,\,.
 \end{eqnarray} 
 And finally $d{\mathcal R}$ in eq.(\ref{kl}) is the volume element of the three-dimensional rotation group and is given by
 \begin{equation}
 d{\mathcal R}\,=\, 4 \, \sin^2{\omega \over{2}}\, d\omega \, \sin{\Theta}\, d\Theta \, d\Phi\,\,.
 \end{equation}

For a statistically isotropic model $ C(\hat{n}_1,\, \hat{n}_2)$ is
invariant under rotation, and therefore $ C({\mathcal R}\hat{n}_1,\,
{\mathcal R}\hat{n}_2)\,=\,C(\hat{n}_1,\, \hat{n}_2)$ and the
orthonormality of $\chi^{\ell}(\omega)$, we will recover the condition
for SI,
 \begin{equation} 
 \kappa^{\ell} \, = \, \kappa^0 \delta_{\ell 0}.
 \end{equation}

Real-space representation of BiPS is very suitable for analytical
computation of BiPS for theoretical models where we know the analytical
expression for the two point correlation of the model, such as
theoretical models in \cite{us_prl}.  On the other hand, the harmonic
representation of BiPS that we describe next allows computationally
rapid methods for BiPS estimation from a given CMB map.

Two point correlation of CMB anisotropies, $C(\hat{n}_1,\,
\hat{n}_2)$, is a two point function on $S^2 \times S^2$, and hence
can be expanded as
\begin{equation}\label{bipolar}
C(\hat{n}_1,\, \hat{n}_2)\, =\, \sum_{l_1,l_2,L,M} A_{l_1l_2}^{\ell M}
\{Y_{l_1}(\hat{n}_1) \otimes Y_{l_2}(\hat{n}_2)\}_{\ell M},
\end{equation}
where $A_{l_1l_2}^{\ell M}$ are coefficients of the expansion (here
after BipoSH coefficients) and $\{Y_{l_1}(\hat{n}_1) \otimes
Y_{l_2}(\hat{n}_2)\}_{\ell M}$ are the bipolar spherical harmonics
which transform as a spherical harmonic with $\ell,\, M$ with respect
to rotations \cite{Var} given by
\begin{equation}
\{Y_{l_1}(\hat{n}_1) \otimes Y_{l_2}(\hat{n}_2)\}_{\ell M} \,=\,
\sum_{m_1m_2} {\mathcal C}_{l_1m_1l_2m_2}^{\ell M} Y_{l_1 m_1}(\hat{n}_2)Y_{l_2 m_2}(\hat{n}_2),
\end{equation}
in which ${\mathcal C}_{l_1m_1l_2m_2}^{\ell M}$ are Clebsch-Gordan
coefficients. We can inverse-transform $C(\hat{n}_1,\, \hat{n}_2)$ to
get the $A_{l_1l_2}^{\ell M}$ by multiplying both sides of
eq.(\ref{bipolar}) by $\{Y_{l'_1}(\hat{n}_1) \otimes
Y_{l'_2}(\hat{n}_2)\}_{\ell'M'}^*$ and integrating over all angles,
then the orthonormality of bipolar harmonics implies that
\begin{equation}\label{alml1l2}
A_{l_1l_2}^{\ell M} \,=\,\int d\Omega_{\hat{n}_1}\int d\Omega_{\hat{n}_2} \,
C(\hat{n}_1,\, \hat{n}_2)\, \{Y_{l_1}(\hat{n}_1) \otimes Y_{l_2}(\hat{n}_2)\}_{\ell M}^*. 
\end{equation}
The above expression and the fact that $C(\hat{n}_1,\, \hat{n}_2)$ is
symmetric under the exchange of $\hat{n}_1$ and $\hat{n}_2$ lead to
the following symmetries of $A_{l_1l_2}^{\ell M}$
\begin{eqnarray}  \label{sym}
A_{l_2l_1}^{\ell M}\,&=&\,(-1)^{(l_1+l_2-L)}A_{l_1l_2}^{\ell M}, \\ \nonumber
A_{ll}^{\ell M} \, &=& \, A_{ll}^{\ell M} \,\,\delta_{\ell,2k+1}, \,\,\,\,\,\,\,\,\,\,\,\,\,\,\,\,\,\,\,\,\, k=0,\,1,\,2,\,\cdots.
\end{eqnarray} 

\begin{figure}[h]
\includegraphics[scale=0.9, angle=0]{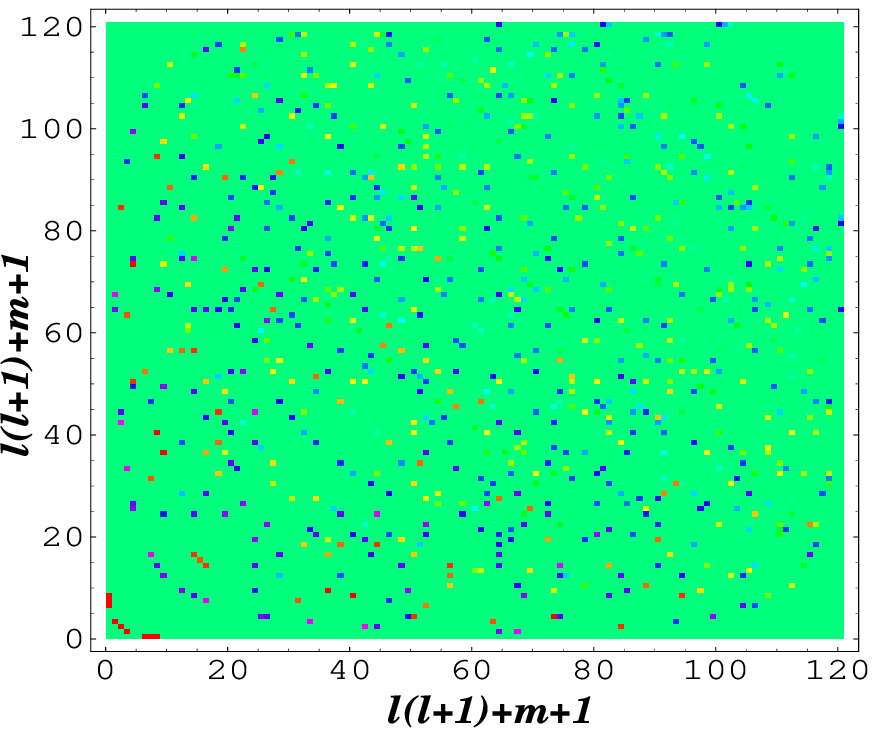}
\includegraphics[scale=0.9, angle=0]{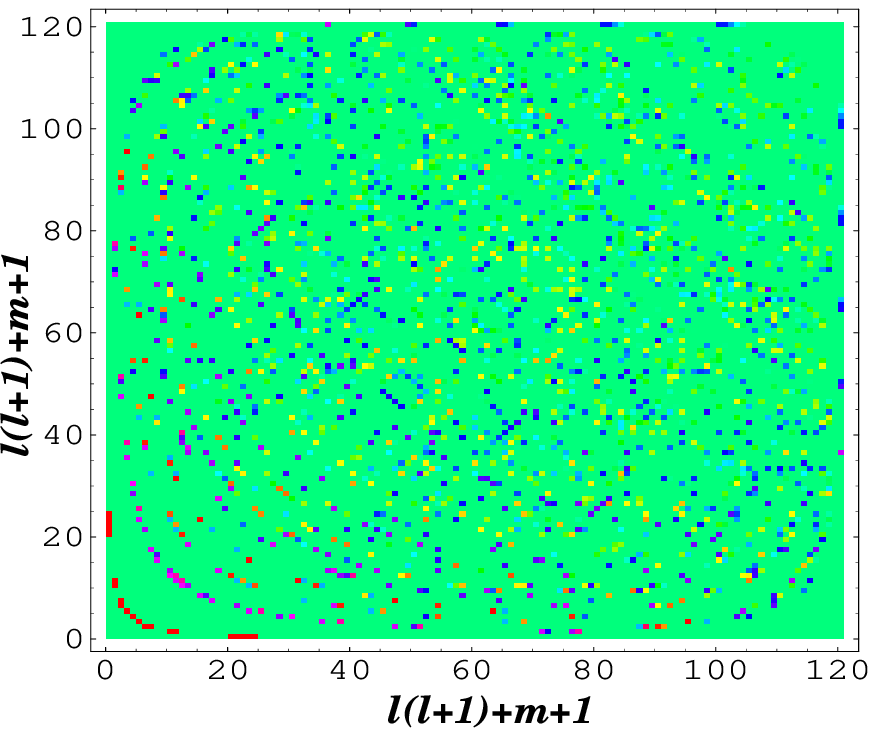}
\caption{BipoSH coefficients are linear combinations of elements of
the covariance matrix. Here $A^{2M}_{ll'}$ ({\it left}) and
$A^{4M}_{ll'}$ ({\it right}) are plotted to show how BiPS covers the
off-diagonal elements of the covariance matrix in harmonic space. }
\label{ALM}
\end{figure}

The Bipolar Spherical Harmonic (BipoSH) coefficients,
$A_{l_1l_2}^{\ell M}$, are linear combinations of off-diagonal
elements of the covariance matrix,
\begin{equation}\label{ALMvsalm}
A^{\ell M}_{l_1 l_2}\,=\, \sum_{m_1m_2} \langle a_{l_1m_1}a^{*}_{l_2 m_2}\rangle (-1)^{m_2} C^{\ell M}_{l_1m_1l_2 -m_2}.
\end{equation}
This means that $A^{\ell M}_{l_1 l_2}$ completely represent the
information of the covariance matrix. Fig. \ref{ALM} shows how $A^{2
M}_{l_1 l_2}$ and $A^{4 M}_{l_1 l_2}$ combine the elements of the
covariance matrix.  When SI holds, the covariance matrix is diagonal
and hence
\begin{eqnarray}  \label{SIALM}
A_{ll^\prime}^{\ell M}\,&=&\,(-1)^l C_{l} (2l+1)^{1/2} \,  \delta_{ll^\prime}\, \delta_{\ell 0}\, \delta_{M0},
\\ \nonumber
A^{0 0}_{l_1 l_2}\,&=&\, (-1)^{l_1} \sqrt{2l_1+1}\, C_{l_1}\, \delta_{l_1l_2}.
\end{eqnarray}  

BipoSH expansion is the most general representation of the two point
correlation functions of CMB anisotropy. The well known angular power
spectrum, $C_l$ is a subspace of BipoSH coefficients corresponding to
the $A_{ll}^{00}$ that represent the statistically isotropic part of a
general correlation function. When SI holds, $A_{ll}^{00}$ or
equivalently $C_l$ have all the information of the field. But if the
SI breaks down, $A_{ll}^{00}$ are not adequate for describing the
field, and one needs to take the other terms into account. This simply
means that the And when the statistical isotropy holds, these
coefficients will reduce to the well-known angular power spectrum of
CMB anisotropy.

It is impossible to measure all $A^{\ell M}_{l_1 l_2}$ individually
because of cosmic variance. Combining BipoSH coefficients into Bipolar
Power Spectrum reduces the cosmic variance\footnote{This is similar to
combining $a_{lm}$ to construct the angular power spectrum,
$C_l=\frac{1}{2l+1}\sum_{m}{|a_{lm}|^2}$, to reduce the cosmic
variance}. BiPS of CMB anisotropy is defined as a convenient
contraction of the BipoSH coefficients 
\begin{equation}\label{kappal}
\kappa_\ell \,=\, \sum_{l,l',M} |A_{ll'}^{\ell M}|^2 \geq 0.
\end{equation} 
The BiPS, which can be shown that is equivalent to the one in
eq.(\ref{kl}), has interesting properties. It is orientation
independent and is invariant under rotations of the sky. For models in
which statistical isotropy is valid, BipoSH coefficients are given by
eq.~(\ref{SIALM}). And results in a null BiPS, {\it i.e.}
$\kappa_\ell\,=\,0$ for every positive $\ell$,
\begin{equation}\kappa_\ell\,=\,\kappa_0 \delta_{\ell 0}.
\end{equation}

\subsection{Unbiased Estimator of BiPS}

An estimator for measuring BipoSH coefficients from a given CMB map is
\begin{equation}\label{estimator}
\tilde A_{ll^\prime}^{\ell M} = \sum_{m m^\prime} \sqrt{W_l W_{l'}} a_{lm}a_{l^\prime
m^\prime} \, \, {\mathcal{ C}}^{\ell M}_{lml^\prime m^\prime}\,\quad ,
\end{equation}
where $W_l$ is the Legendre transform of the window function. The
above estimator is a linear combination of $C_l$ and hence is
unbiased. An unbiased estimator of BiPS is given by
\begin{equation} 
\tilde\kappa_\ell = \sum_{ll^\prime M} \left|\tilde A_{ll^\prime}^{\ell
M}\right|^2 - {\mathfrak B}_\ell\, ,
\end{equation} 
where the bias for the BiPS is defined as ${\mathfrak
B}_\ell=\langle\tilde\kappa_\ell\rangle-\kappa_\ell$ is equal to
\begin{eqnarray}\label{klbias}
{\mathfrak B}_\ell = \sum_{l_1,l_2}W_{l_1}
W_{l_2}\,\, \sum_{m_1,m_1^\prime} \sum_{m_2,m_2^\prime}&&\,\left[\langle
a^*_{l_1m_1}a_{l_1 m_1^\prime}\rangle\langle a^*_{l_2m_2}a_{l_2
m_2^\prime}\rangle + \langle a^*_{l_1m_1}a_{l_2
m_2^\prime}\rangle\langle a^*_{l_2m_2}a_{l_1 m_1^\prime}\rangle
\right] \nonumber \\ &&{}\times \sum_M \mathcal{ C}^{\ell
M}_{l_1m_1l_2m_2}\mathcal{ C}^{\ell M}_{l_1m_1^\prime
l_2m_2^\prime}\,.
\end{eqnarray}

\begin{figure}[h]
 \includegraphics[scale=0.3, angle=-90]{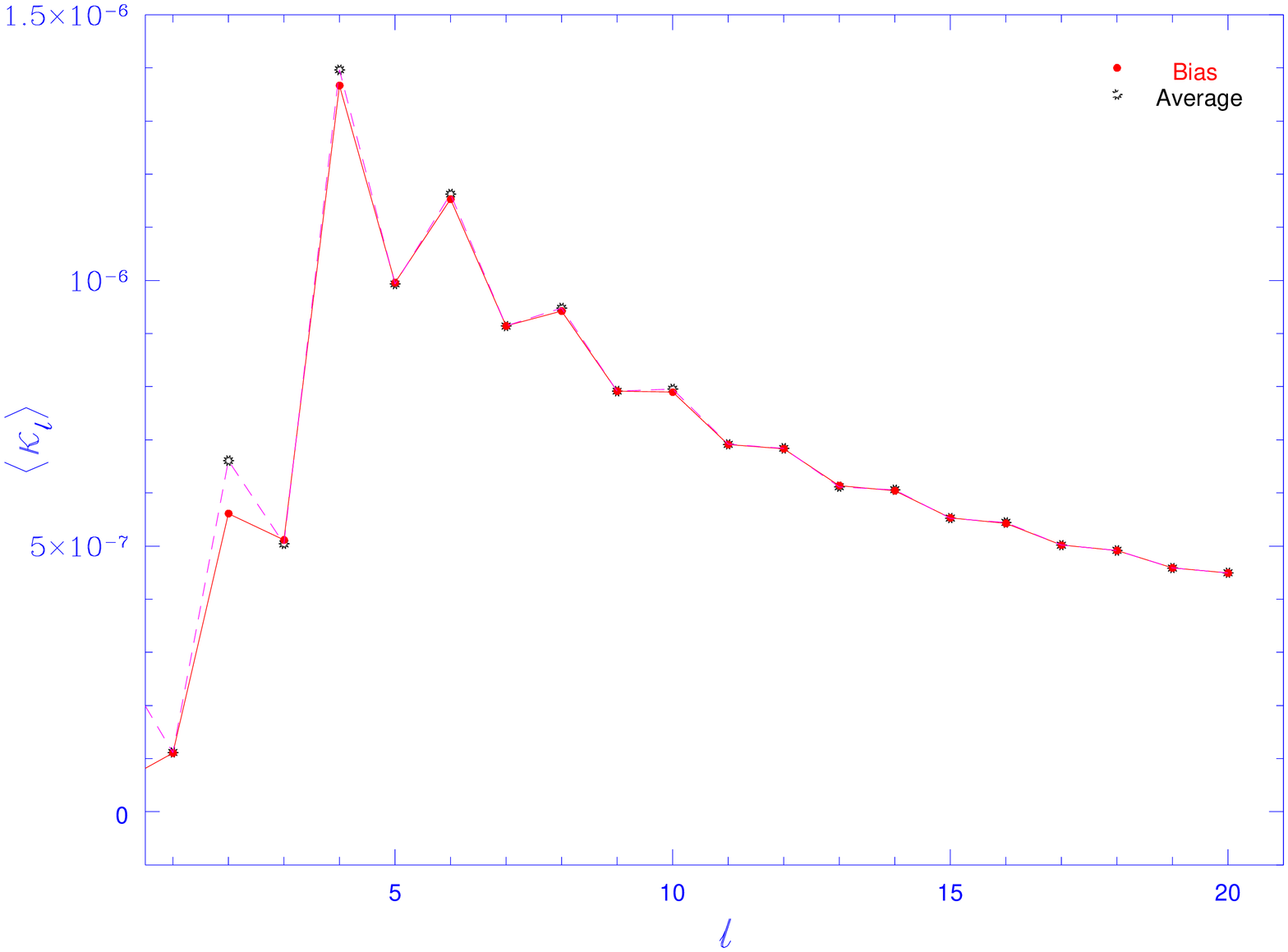}
 \includegraphics[scale=0.3, angle=-90]{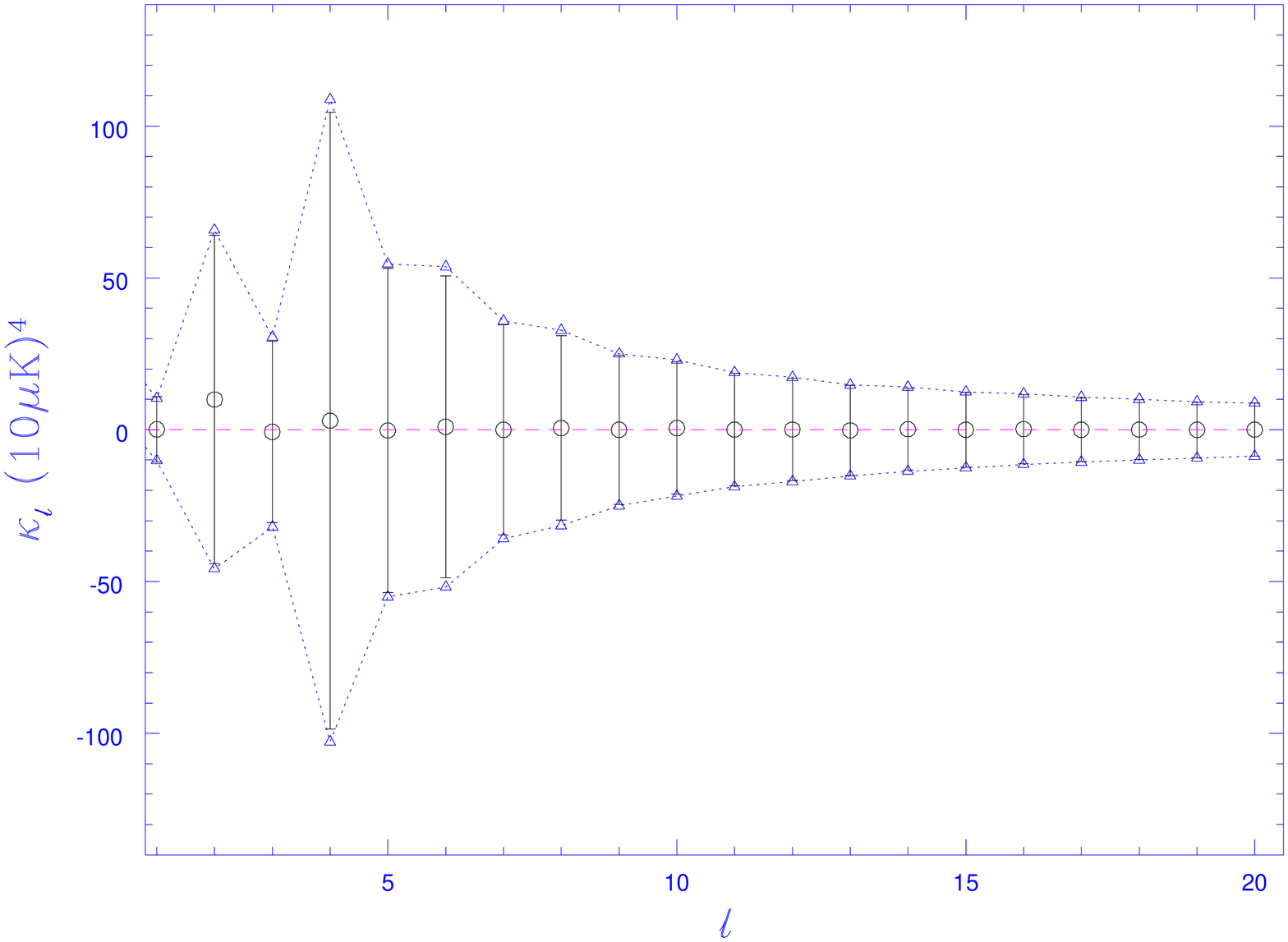} 
  \caption{{\bf Left:} Analytical bias for a Gaussian window function
  with $W_l^G(40)$ computed from the average $C_l$ from 1000
  realizations of a SI CMB map compared with $\langle
  \kappa_l^{realization}\rangle$ (the average $\kappa_l$ from 1000
  realizations). This shows that the theoretical bias is a very good
  estimation of the bias for a statistically isotropic map. {\bf
  Right:} The cosmic error, $\sigma({\kappa_\ell})$, obtained using
  $1000$ independent realizations of CMB (full) sky map matches the
  analytical results shown by dotted curve with triangles . This shows
  a much better fit to the theoretical cosmic variance compared to
  what was obtained for $100$ realizations \cite{us_apjl} }
  \label{bias3020}
 \label{comparison3020}
\end{figure}

The above expression for ${\mathfrak B}_\ell$ is obtained by assuming
Gaussian statistics of the temperature fluctuations. The procedure is
very similar to computing cosmic variance (which is discussed in the
next section), but much simpler. However, we can not measure the
ensemble average in the above expression and as a result, elements of
the covariance matrix (obtained from a single map) are poorly
determined due to the cosmic variance. The best we can do is to
compute the bias for the SI component of a map
\begin{equation}\label{klisobias}
{\mathfrak B}_\ell \equiv\langle\tilde\kappa_\ell^B\rangle_{_{\rm SI}}
 = (2\ell+1)\,\sum_{l_1} \sum_{l_2=|\ell-l_1|}^{\ell+l_1} C_{l_1}
 C_{l_2} W_{l_1} W_{l_2} \left[1 + (-1)^{\ell}\, \delta_{l_1
 l_2}\right]\,.
\end{equation}
{\em Note , the estimator $\tilde \kappa_\ell$ is unbiased, only for
SI correlation,i.e., $\langle \tilde \kappa_\ell \rangle=0$.}
Consequently, for SI correlation, the measured $\tilde \kappa_\ell$
will be consistent with zero within the error bars given by
$\sigma_{_{\rm SI}}$ \cite{us_apjl}.  We simulated 1000 SI CMB maps
and computed BiPS for them using different filters. The average BiPS
of SI maps is an estimation of the bias which can be compared to our
analytical estimation. The left panel of Fig. \ref{bias3020} shows
that the theoretical bias (computed from average $C_l$) match the
numerical estimations of average $\kappa_{\ell}$ of the 1000
realizations of the SI maps.

It is important to note that bias cannot be correctly subtracted for
non-SI maps. Non-zero $\tilde \kappa_\ell$ estimated from a non-SI map
will have contribution from the non-SI terms in full bias given in
eq.~(\ref{klbias}). It is not inconceivable that for strong SI
violation, ${\mathfrak B}_\ell$ over-corrects for the bias leading to
negative values of $\tilde \kappa_\ell$. {\em What is important is
whether measured $\tilde \kappa_\ell$ differs from zero at a
statistically significant level.}

\subsection{Cosmic Variance of BiPS}

A crucial point is how well one can hope to estimate the BiPS given
the single observed sky. This is limited by the Cosmic variance of the
BiPS estimator defined as
\begin{equation} 
\sigma^2\,=\,<\tilde{\kappa}_{\ell}^2>-<\tilde{\kappa}_{\ell}>^2
\end{equation}
It is possible to obtain an analytic expression variance of
$\tilde{\kappa}^{\ell}$ using the Gaussianity of $\Delta T$. Looking
back at the eq.(\ref{kl}) we can see, we will have to calculate the
eighth moment of the field
\begin{equation} 
\langle \Delta T(\hat{n}_1)\Delta T(\hat{n}_2)\Delta T(\hat{n}_3)\Delta T(\hat{n}_4)\Delta
T(\hat{n}_5)\Delta T(\hat{n}_6)\Delta T(\hat{n}_7)\Delta T(\hat{n}_8) \rangle.
\end{equation}
Assuming Gaussianity of the field we can rewrite the eight point
correlation in terms of two point correlations.  One can write a
simple code to do that{\footnote{F90 software implementing this is
available from the authors upon request}}. This will give us
$(8-1)!!=7\times 5 \times 3=105$ terms. These 105 terms consist of
terms like:
\begin{equation} 
\langle \Delta T(\hat{n}_1)\Delta T(\hat{n}_2) \rangle\langle\Delta
T(\hat{n}_3)\Delta T(\hat{n}_4) \rangle\langle\Delta
T(\hat{n}_5)\Delta T(\hat{n}_6) \rangle\langle\Delta
T(\hat{n}_7)\Delta T(\hat{n}_8) \rangle,
\end{equation}
and all other permutations of them.  On the other hand
$\langle\tilde{\kappa}_{\ell}\rangle$ has a 4 point correlation in it
which can also be expanded versus two point correlation functions.  If
we form
$\langle\tilde{\kappa}_{\ell}^2\rangle-\langle\tilde{\kappa}_{\ell}\rangle^2$,
only 96 terms will be left which are in the following form
\begin{equation} 
\label{firstterm}
(\frac{2\ell+1}{8\pi^2})^2 \int d\Omega_1 \cdots d\Omega_4 
\int dR \int dR' \chi_{\ell}(R) \chi_{\ell}(R')  C(\hat{n}_1,R'\hat{n}_4)C(\hat{n}_2,R'\hat{n}_3)C(R\hat{n}_1,\hat{n}_4)C(R\hat{n}_2,\hat{n}_3)
\end{equation}
and all other permutations. As described in detail in our paper
~\cite{haj_sour05b}, it is possible to simplify and group together the
$96$ terms and obtain a compact expression as 

\begin{eqnarray}\label{klcv}
&&\sigma^2_{_{\rm SI}}(\tilde\kappa_\ell) =\sum_{l : 2l \ge \ell}\!\!
4\, C_{l}^4 W_l^4 \left[ 2 \frac{(2\ell+1)^2}{2l+1}+ (-1)^{\ell}
(2\ell+1)+ (1+2(-1)^{\ell}) F_{ll}^\ell\right] \nonumber \\
&&{}+\sum_{l_1} \sum_{l_2=|\ell-l_1|}^{\ell+l_1} \!\!\! 4
\,C_{l_1}^2\,C_{l_2}^2 W_{l_1}^2\,W_{l_2}^2\left[ (2\ell+1) +
F_{l_1l_2}^\ell \right]\nonumber \\
&& + \,8\sum_{l_1}\,\frac{(2\ell+1)^2}{2l_1+1}
\,C_{l_1}^2 W_{l_1}^2 \left[\sum_{l_2=|\ell-l_1|}^{\ell+l_1} C_{l_2}
W_{l_2}\right]^2\nonumber \\
&&{} + 16\,(-1)^{\ell}\,\sum_{l_1 : 2l_1 \ge \ell}
\,\frac{(2\ell+1)^2}{2l_1+1}\, \sum_{l_2=|\ell-l_1|}^{\ell+l_1}
C_{l_1}^3 C_{l_2}\,W_{l_1}^3 W_{l_2}.
\end{eqnarray}

Numerical computation of $\sigma^2_{_{\rm SI}}$ is fast. But the
challenge is to compute Clebsch-Gordan coefficients for large quantum
numbers. We use drc3j subroutine of
netlib\footnote{http://www.netlib.org/slatec/src/} in order to compute
the Clebsch-Gordan coefficients in our codes. Again we can check the
accuracy of our analytical estimation of cosmic variance by comparing
it against the standard deviation of BiPS of 1000 simulations of SI
CMB sky. The result is shown the right panel of in
Fig.~\ref{comparison3020} and shows a very good agreement between the
two.

\section{Results of BiPS analysis of WMAP CMB maps} 

We carry out measurement of the BiPS, on the following CMB anisotropy
maps
\begin{itemize}
\item[A)] a foreground cleaned map (denoted as `TOH')~\cite{maxwmap},
\item[B)] the Internal Linear Combination map (denoted as `ILC' in the figures)~\cite{wmap}, and
\item[C)] a customized linear combination of the QVW maps of WMAP with a galactic cut (denoted as `CSSK').
\end{itemize}
Also for comparison, we measure the BiPS of 
\begin{itemize}
\item[D)] a Wiener filtered map of WMAP data (denoted as `Wiener')~\cite{maxwmap}, and
\item[E)] the ILC map with a $10^{\circ}$ cut around the equator (denoted as `Gal. cut.').
\end{itemize}

\begin{figure}[h]
  \includegraphics[scale=0.6, angle=-90]{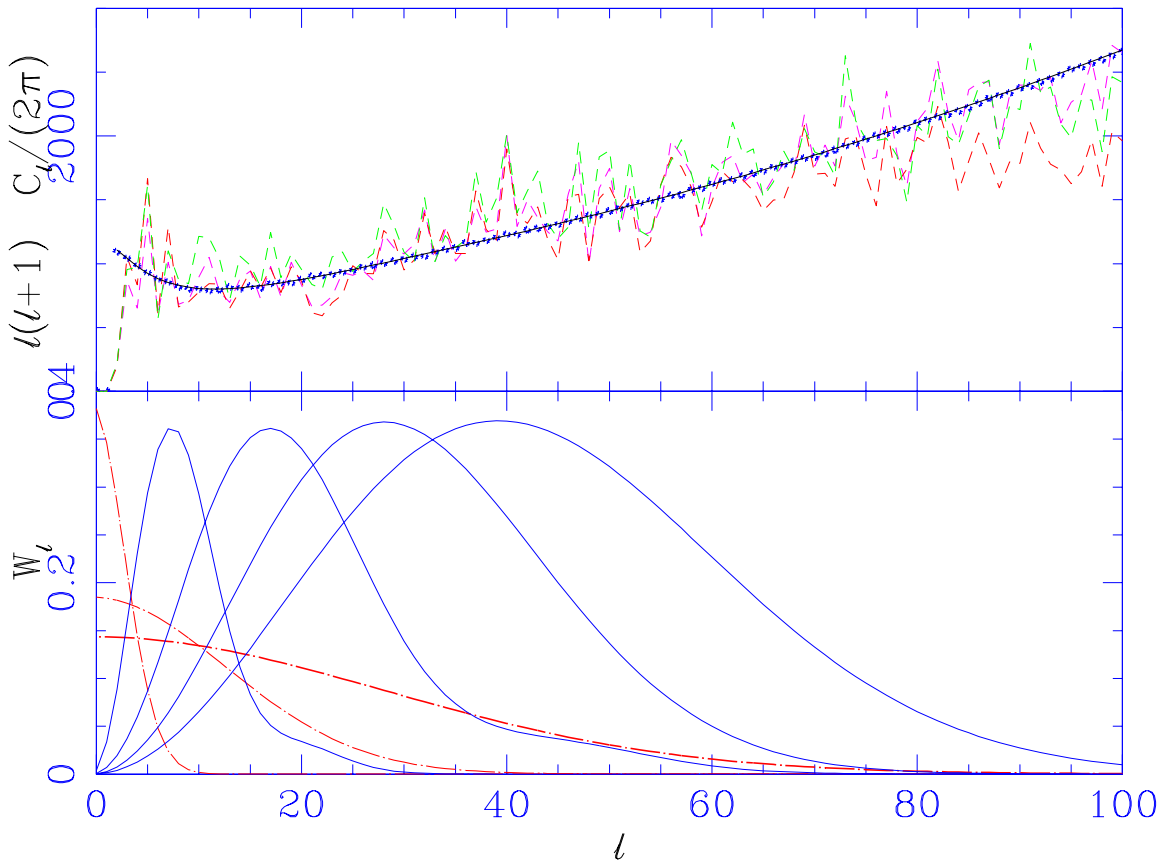}
 \caption{  {\em Top:} 
$C_{\ell}$ of the two WMAP CMB anisotropy maps. The red, magenta and green curves correspond to map A, B and C, respectively. The black line is a `best fit' WMAP theoretical $C_{\ell}$ used for simulating SI maps. Blue dots are the average $C_l$ recovered from $1000$ realizations. {\em Bottom:} These plots show
 the window functions used. The dashed curves with increasing $l$ coverage are `low-pass' filter, $W_l^G(l_s)$, with $l_s=4, 18, 40$, respectively.  The solid lines are `band-pass' filter $W^S_l(l_t,l_s)$ with $(l_s,l_t)=(13,2), (30,5), (30,20), (45,20)$, respectively.}
\label{cl_wmap_cl100}
\end{figure}

Angular power spectra of these maps are shown in
Fig.~\ref{cl_wmap_cl100}. The best fit theoretical power spectrum from
the WMAP analysis~\footnote{Based on an LCDM model with a
scale-dependent (running) spectral index which best fits the dataset
comprised of WMAP, CBI and ACBAR CMB data combined with 2dF and
Ly-$\alpha$ data} ~\cite{sper_wmap03} is plotted on the same figure.
$C_l$ from observed maps are consistent with the theoretical curve,
$C_l^T$ (except for the lowest multipoles). The bias and cosmic
variance of BiPS depend on the total SI angular power spectrum of the
signal and noise $C_l = C_l^S + C_l^N$.  However, we have restricted
our analysis to $l \lsim 60$ where the errors in the WMAP power
spectrum is dominated by cosmic variance.  It is conceivable that the
SI violation is limited to particular range of angular scales.  Hence,
multipole space windows that weigh down the contribution from the SI
region of multipole space will enhance the signal relative to cosmic
error, $\sigma_{_{\rm SI}}$. We use simple filter functions in $l$
space to isolate different ranges of angular scales; a low pass,
Gaussian filter
\begin{equation}
W^G_l(l_s) = \exp(-(l+1/2)^2/(l_s+1/2)^2)
\label{gaussfilter}
\end{equation}
that cuts off power on small angular scales ($\lsim 1/l_s$) and a band
pass filter,
\begin{equation}
W^S_l(l_t, l_s) = \left[2(1- J_0((l+1/2)/(l_t+1/2)))
\right]\,\exp(-(l+1/2)^2/(l_s+1/2)^2)
\label{bpfilter}
\end{equation}
that retains power within a range of multipoles set by $l_t$ and
$l_s$.  The windows are normalized such that $\sum_l (l+1/2)/(l(l+1))
W_l =1$, i.e., unit {\it rms} for unit flat band power
$C_l=1/(l(l+1))$. The window functions used in our work are plotted in
figure~\ref{cl_wmap_cl100}.  We use the $C_l^T$ to generate 1000
simulations of the SI CMB maps. $a_{lm}$'s are generated up to an
$l_{max}$ of 1024 (corresponding to HEALPix resolution
$N_{side}=512$). These are then multiplied by the window functions
$W^G_l(l_s)$ and $W^S_l(l_t, l_s)$. We compute the BiPS for each
realization.  Fig.\ref{cl_wmap_cl100} shows that the average power
spectrum obtained from the simulation matches the theoretical power
spectrum, $C_l^T$, used to generate the realizations. We use $C_l^T$
to analytically compute bias and cosmic variance estimation for
$\tilde{\kappa_{\ell}}$. This allows us to rapidly compute BiPS with
$1\sigma$ error bars for different theoretical $C_l^T$.

\begin{figure}[h]
  \includegraphics[scale=0.6, angle=-90]{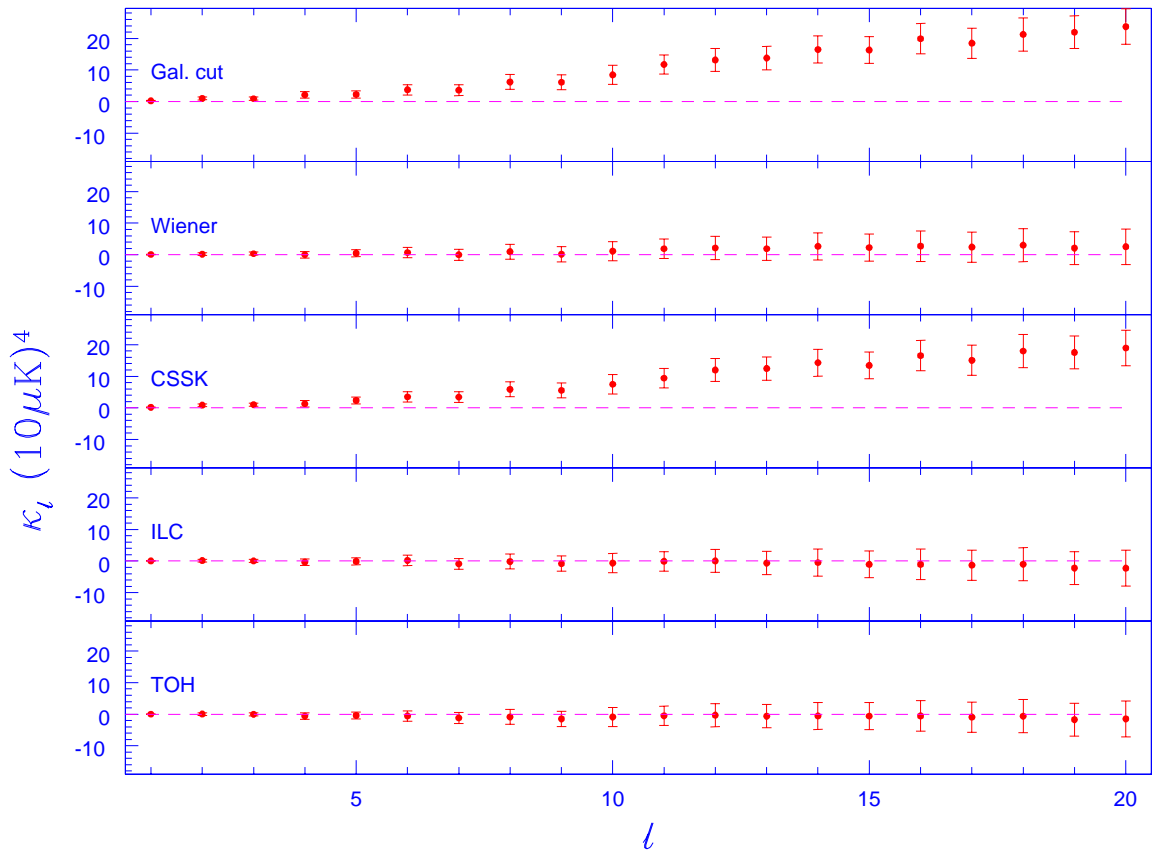}
  \caption{Measured BiPS for maps A, B and C filtered with a window
with $l_s =30$, $l_t =20$. This is to check the statistical isotropy
of the WMAP in the modest $20<l<40$ range in the multipole space where
certain anomalies have been reported. ILC with a 10-degree-cut (top)
has the same BiPS as map C ($l_s =30$, $l_t =20$) which explains that
the raising tail of CSSK map is because of the mask. }
  \label{kappa_wmap_3020}
\end{figure}

We use the estimator given in eq.(\ref{estimator}) to measure BiPS
for the given CMB maps. We compute the BiPS for all window functions
shown in Fig~\ref{cl_wmap_cl100}. Results for one these windows are
plotted in Figs. \ref{kappa_wmap_3020}. In the low-$l$ regime, where
we have kept the low multipoles, BiPS for all three given maps are
consistent with zero. But in the intermediate-$l$ regime
(Fig. \ref{kappa_wmap_3020}), although BiPS of ILC and TOH maps are
well consistent with zero, the CSSK map shows a rising tail in BiPS
due to the galactic mask. To confirm it, we compute the BiPS for the
ILC map with a 10-degree cut around the galactic plane (filtered with
the same window function). The result is shown on the top panel of
Fig. \ref{kappa_wmap_3020}. Another interesting effect is seen when we
apply a $W_l^S(20,45)$ filter, where Wiener filtered map has a non
zero BiPS very similar to that of CSSK but weaker. The reason is that
Wiener filter takes out more modes from regions with more foregrounds
since these are inconsistent with the theoretical model. As a result,
a Wiener filtered map at $W_l^s(20,45)$ filter has a BiPS similar to a
cut sky map. The fact that Wiener map has less power at the Galactic
plane can even be seen by eye! Hence using different filters allows us
to uncover different types of violation of SI in a CMB map. In our
analysis we have used a set of filters which enables us to probe SI
breakdown on angular scales $l\lsim 60$.

\begin{figure}[h]
\begin{center}
  \includegraphics[scale=0.6, angle=0]{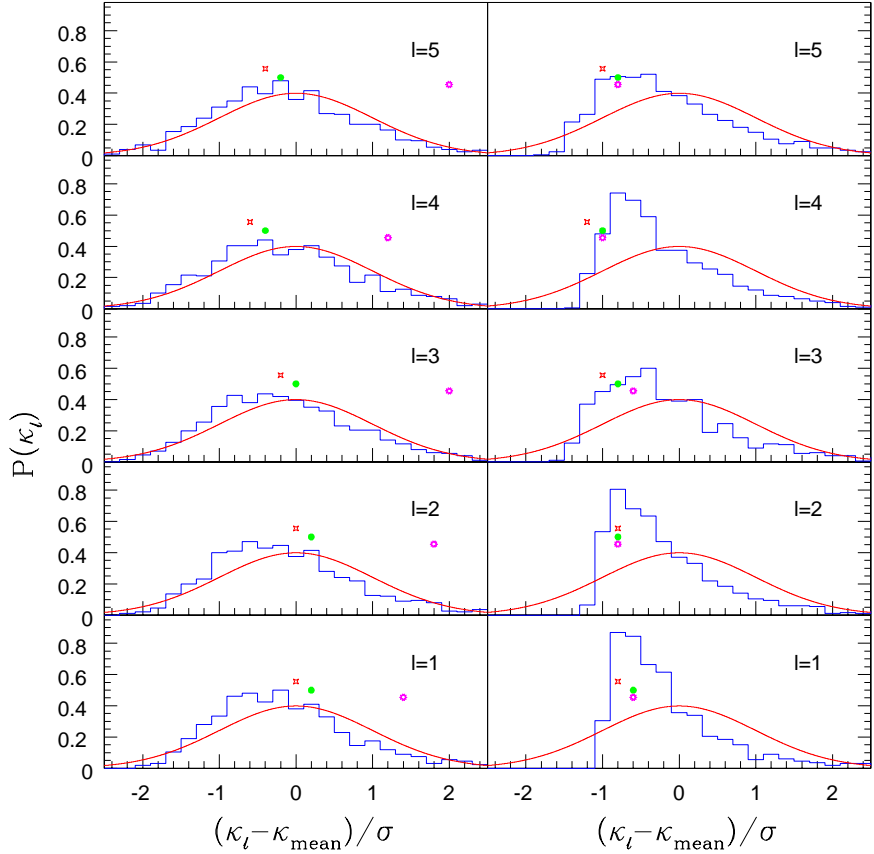}
  \caption{ Probability distribution function for $\kappa_1$ to
$\kappa_5$ constructed from $1000$ realizations. The left panel shows
the PDF for the maps filtered with $W^S_l(20,30)$ (left panel) and
$W^G_l(l_s=40)$ (right panel). The latter is more skewed, which
explains the apparent $\sim 1\sigma$ shift in the $\kappa_{\ell}$
values for $W^G_l(l_s=40)$ at low $\ell$. The green, magenta and red
(circular, pentagonal and rectangular) points represent ILC, CSSK and
TOH maps, respectively. The smooth solid curves are Gaussian
approximations.}
\label{pdf1to5}
\end{center}
\end{figure}

\begin{figure}[h]
\begin{center}
  \includegraphics[scale=0.6, angle=0]{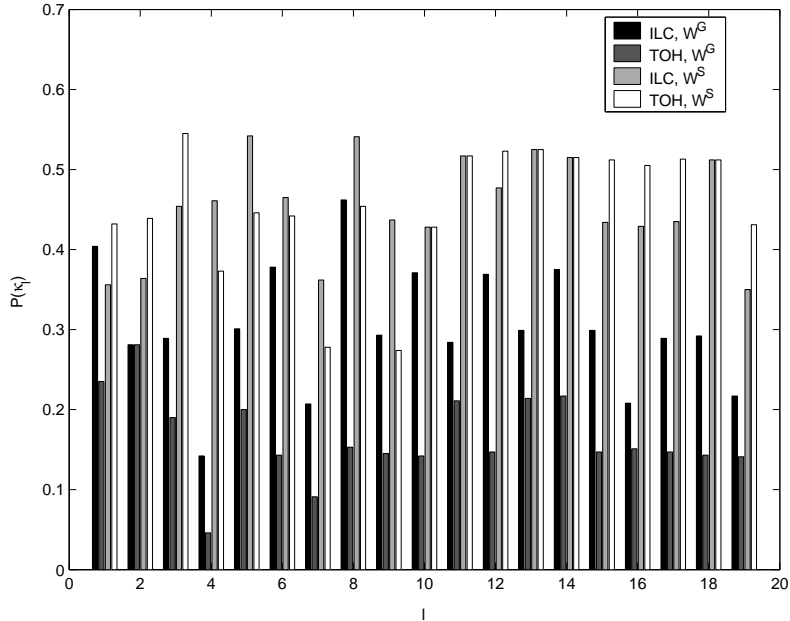}
  \caption{The probability of two of the {\it WMAP} based CMB maps
being SI when filtered by $W_l^S(20,30)$ and a Gaussian filter
$W_l^G(40)$. }
\end {center}
\label{prob3020}
\label{prob400}
\end{figure}

The BiPS measured from $1000$ simulated SI realizations of $C_l^T$ is
used to estimate the probability distribution functions (PDF),
$p(\tilde \kappa_{\ell})$. A sample of the PDF for two windows is
shown in Fig.~\ref{pdf1to5}. Measured values of BiPS for ILC, TOH and
CSSK maps are plotted on the same plot. BiPS for ILC and TOH maps are
located very close to the peak of the PDF.  We compute the individual
probabilities of the map being SI for each of the measured
$\tilde{\kappa}_{\ell}$. This probability is obtained by integrating
the PDF beyond the measured $\tilde \kappa_{\ell}$. To be precise, we
compute
\begin{eqnarray} 
\label{probability}
P(\tilde\kappa_{\ell}| {C_l^T})&=& P(\kappa_{\ell} >
\tilde\kappa_{\ell})=\int_{\tilde\kappa_{\ell}}^{\infty} d\kappa_{\ell}
\,p(\kappa_{\ell}), \,\,\,\,\,\tilde\kappa_{\ell} > 0, \\ \nonumber &=&
P(\kappa_{\ell} < \tilde\kappa_{\ell})=\int_{-\infty}^{\tilde\kappa_{\ell}}d\kappa_{\ell}
\,p(\kappa_{\ell}), \,\,\,\,\,\tilde\kappa_{\ell} < 0.  
\end{eqnarray}  
The probabilities obtained are shown in Figs. \ref{prob3020} and
\ref{prob400} for $W^S(20,30), W_l^G(40)$ and $W_l^G(4)$. The
probabilities for the $W^S_l(20,30)$ window function are greater than
$0.25$ and the minimum probability at $\sim 0.05$ occurs at $\kappa_4$
for $W^G(40)$. The reason for systematically lower SI probabilities
for $W{_l}^S(20,30)$ as compared to $W{_l}^G(40)$ is simply due to
lower cosmic variance of the former. The contribution to the cosmic
variance of BiPS is dominated by the low spherical harmonic
multipoles. Filters that suppress the $a_{lm}$ at low multipoles have
a lower cosmic variance.

It is important to note that the above probability is a conditional
probability of measured $\tilde\kappa_{\ell}$ being SI given the
theoretical spectrum $C_l^T$ (used to estimate the bias).  A final
probability emerges as the Bayesian chain product with the probability
of the theoretical $C_l^T$ used given data. Hence, small difference in
these conditional probabilities for the two maps are perhaps not
necessarily significant. Since the BiPS is close to zero, the
computation of a probability marginalized over the $C_l^T$ may be
possible using Gaussian (or, improved) approximation to the PDF of
$\kappa_{\ell}$. 

\begin{figure}[h]
  \includegraphics[scale=0.3, angle=-90]{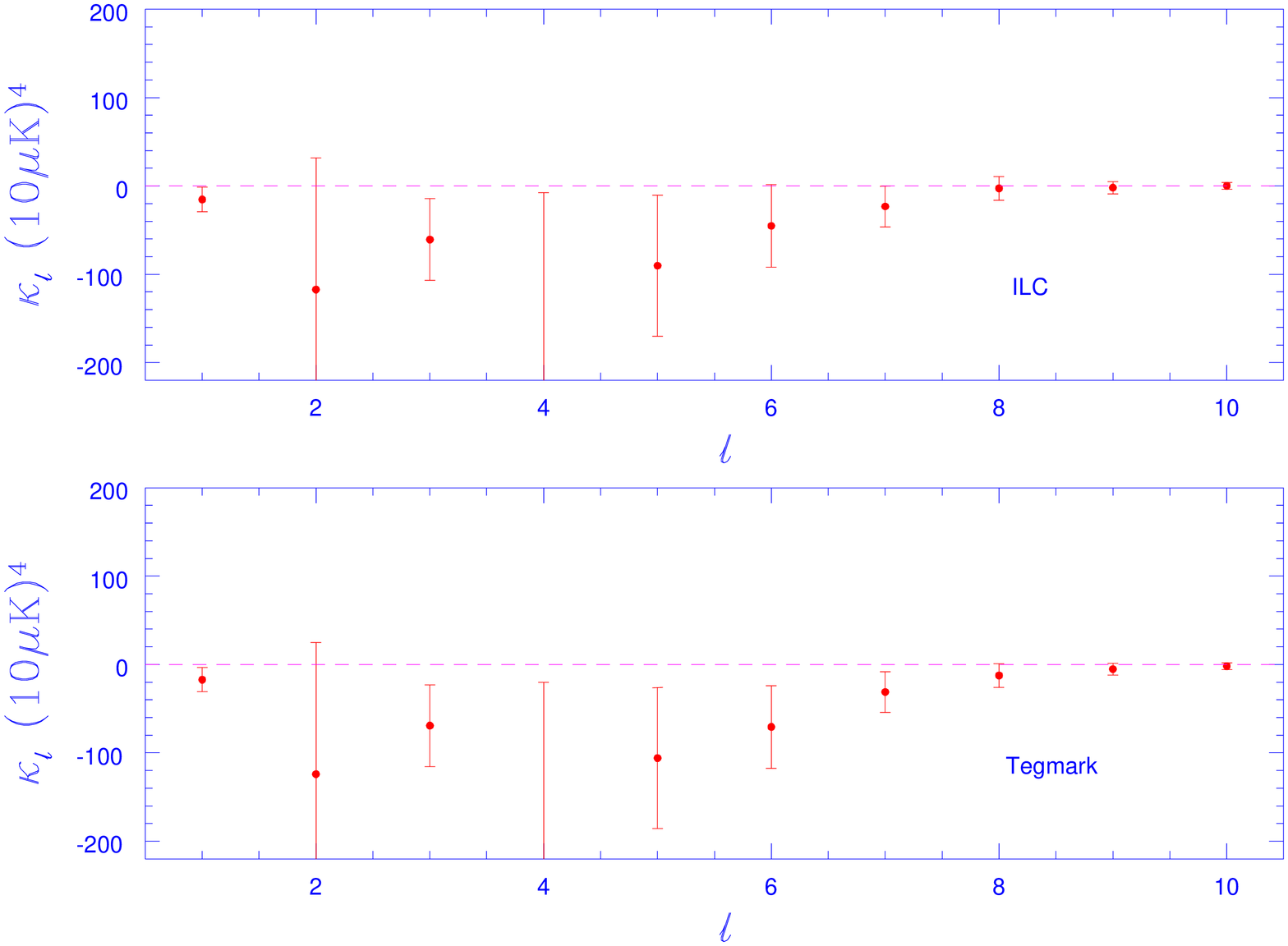}
  \includegraphics[scale=0.3, angle=-90]{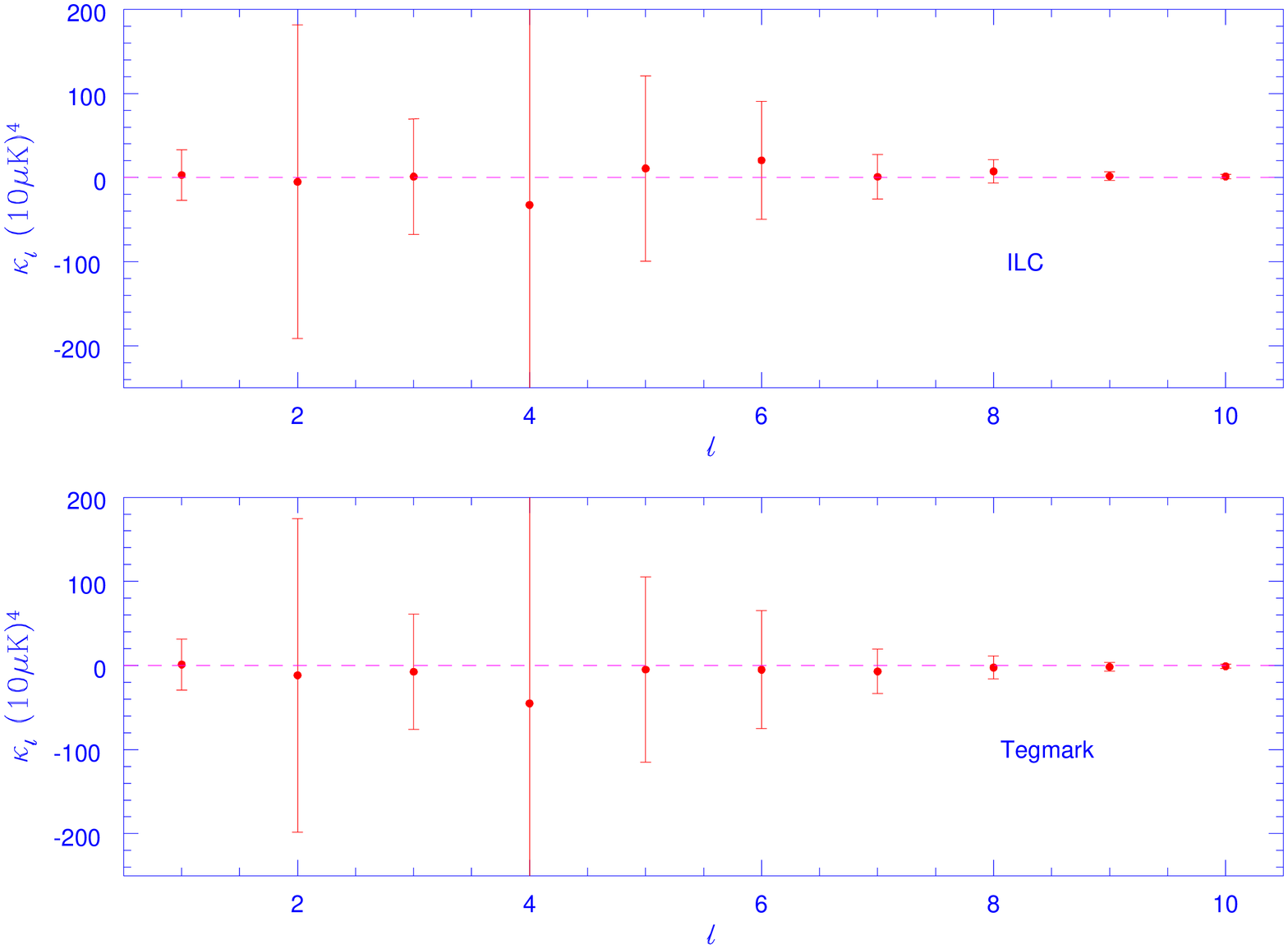}
  \includegraphics[scale=0.45, angle=0]{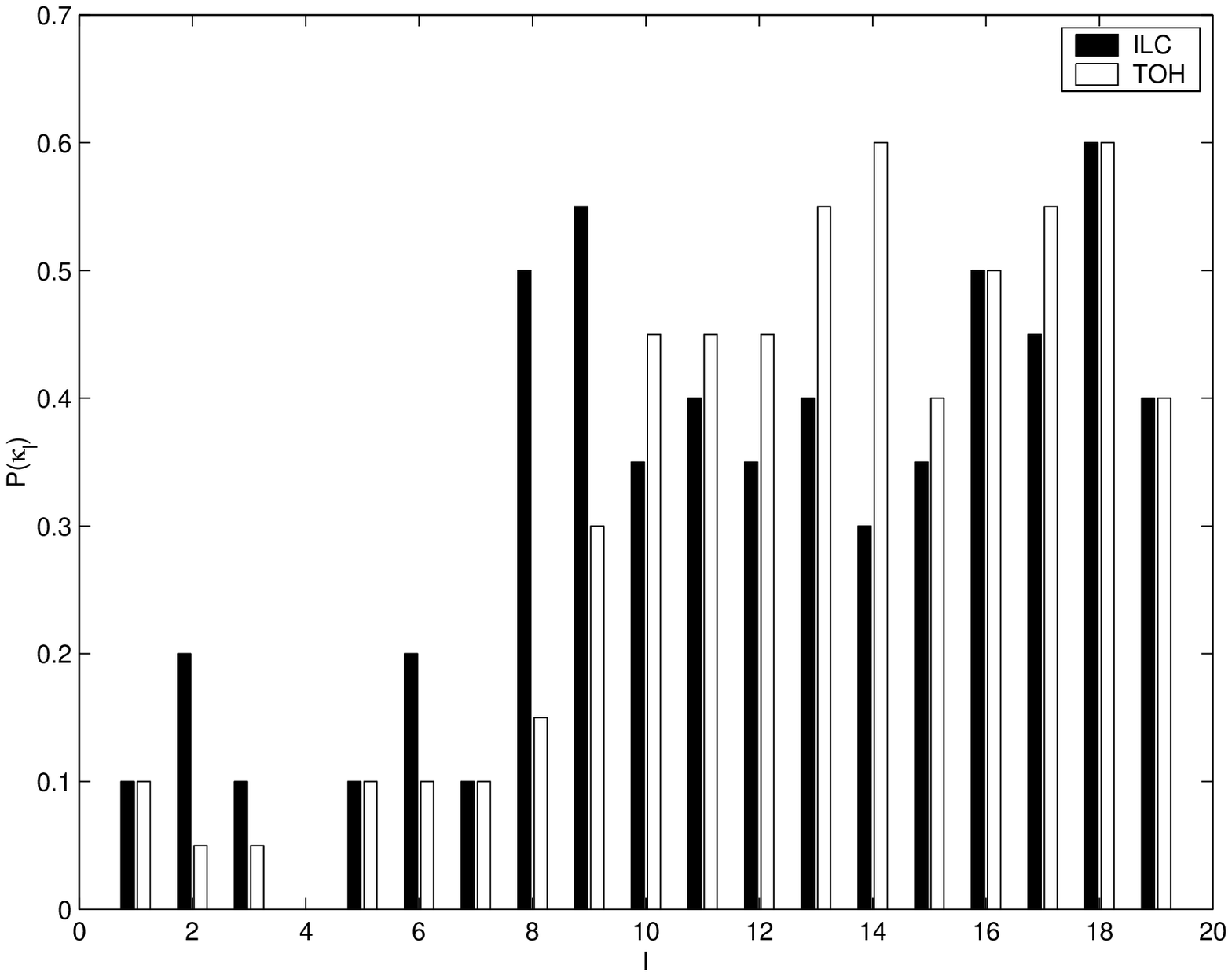}
  \includegraphics[scale=0.45, angle=0]{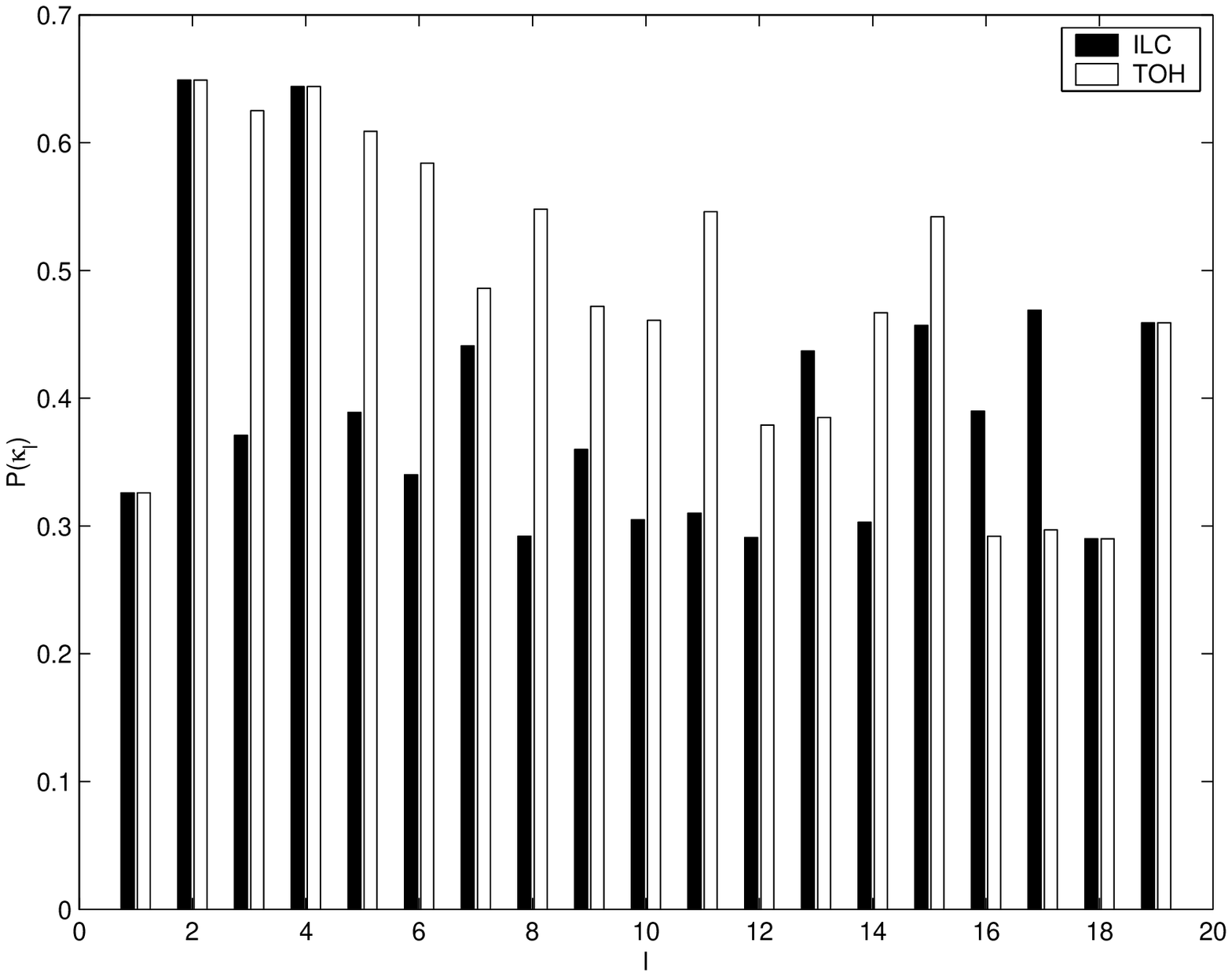}
  \caption{{\bf Top;} Figure compares the measured values of
$\kappa_\ell$ for maps A and B filtered to retain power only on the
lowest multipoles, $l=2$ and $l=3$ assuming the WMAP theoretical
spectrum WMAPbf ({\it left}) and a model spectrum that matches the
suppressed power at the lowest multipoles \cite{shaf_sour04}. The non
zero $\kappa_{\ell}$ `detections' assuming the WMAP theoretical
spectrum become consistent with zero for a $C_l^T$ that has power
suppressed at low multipoles.  {\bf Bottom:} The corresponding SI
probability assuming the WMAP theoretical spectrum, $C_l^T$
({\it{left}}) and a model spectrum that matches the suppressed power
at the lowest multipoles ({\it right}).  }

\label{kappa_wmap_0_4}
\label{prob40}
\end{figure}
The important role played by the choice of the theoretical model for
the BiPS measurement is shown for a $W_l$ that retains power in the
lowest multipoles, $l=2$ and $l=3$. Assuming $C_l^T$, there are hints
of non-SI detections in the low $\ell$'s (top-left panel of
Fig.~\ref{kappa_wmap_0_4}). We also compute the BiPS using a $C_l^T$
for a model that accounts for suppressed quadrupole and octopole in
the WMAP data~\cite{shaf_sour04}. The mild detections of a non zero
BiPS vanish for this case (top-right panel of
Fig.~\ref{kappa_wmap_0_4}).  The corresponding SI probabilities for
the two choices of $C_l^T$ are shown in the lower panels.

\section{Discussion and Conclusion}

The SI of the CMB anisotropy has been under scrutiny after the release
of the first year of WMAP data. We use the BiPS which is sensitive to
structures and patterns in the underlying total two-point correlation
function as a statistical tool of searching for departures from SI.
We carry out a BiPS analysis of WMAP full sky maps. We find no strong
evidence for SI violation in the WMAP CMB anisotropy maps considered
here. We have verified that our null results are consistent with
measurements on simulated SI maps.  The BiPS measurement reported here
is a Bayesian estimate of the conditional probability of SI (for each
$\kappa_{\ell}$ of the BiPS) given an underlying theoretical spectrum
$C_l^T$. We point out that the excess power in the $C_l^T$ with
respect to the measured $C_l$ from WMAP at the lowest multipoles tends
to indicate mild deviations from SI. BiPS measurements are shown to be
consistent with SI assuming an alternate model $C_l^T$ that is
consistent with suppressed power on low multipoles. Note that it is
possible to band together $\kappa_{\ell}$ measurements to tighten the
error bars further. The full sky maps and the restriction to low
$l<60$ (where instrumental noise is sub-dominant) permits the use of
our analytical bias subtraction and error estimates. The excellent
match with the results from numerical simulations is a strong
verification of the numerical technique.  This is an important check
before using Monte-Carlo simulations in future work for computing BiPS
from CMB anisotropy sky maps with a galactic mask and non uniform
noise matrix.

There are strong theoretical motivations for hunting for SI violation
in the CMB anisotropy. The possibility of non-trivial cosmic topology
is a theoretically well motivated possibility that has also been
observationally targeted~\cite{ell71, lac_lum95, lev02, linde}.  The
breakdown of statistical homogeneity and isotropy of cosmic
perturbations is a generic feature of ultra large scale structure of
the cosmos, in particular, of non trivial cosmic topology \cite{bps}.
The underlying correlation patterns in the CMB anisotropy in a
multiply connected universe is related to the symmetry of the
Dirichlet domain. The BiPS expected in flat, toroidal models of the
universe has been computed and shown to be related to the principle
directions in the Dirichlet domain \cite{us_prl}. As a tool for
constraining cosmic topology, the BiPS has the advantage of being
independent of the overall orientation of the Dirichlet domain with
respect to the sky. Hence, the null result of BiPS can have important
implication for cosmic topology. This approach complements direct
search for signature of cosmic topology~\cite{circles, staro} and our
results are consistent with the absence of the matched circles and the
null S-map test of the WMAP CMB maps~\cite{circles04, angelwmap}. Full
Bayesian likelihood comparison to the data of specific cosmic topology
models is another approach that has applied to COBE-DMR
data~\cite{bps}. Work is in progress to carry out similar analysis on
the large angle WMAP data.  We defer to future publication, detailed
analyzes and constraints on cosmic topology using null BiPS
measurements, and the comparison to the results from complementary
approaches. There are also other theoretical scenarios that predict
breakdown of SI that can be probed using BiPS, e.g., primordial
cosmological magnetic fields \cite{DKY, gang}.

The null BiPS results also has implications for the observation and
data analysis techniques used to create the CMB anisotropy
maps. Observational artifacts such as non-circular beam, inhomogeneous
noise correlation, residual stripping patterns, etc.  are potential
sources of SI breakdown.  Our null BiPS results confirm that these
artifacts do not significantly contribute to the maps studied
here. Foreground residuals can also be sources of SI breakdown. The
extent to which BiPS probes foreground residuals is yet to be fully
studied and explored. We do not see any significant effect of the
residual foregrounds in ILC and the TOH maps as it was mentioned by
\cite{erik04c}. This can not be necessarily called a discrepancy
between the two results unless we know what should have been seen in
the BiPS. The question is if the signal is strong enough and whether
the effect smeared out in bipolar multipole space within our angular
$l$-space window. On the other hand, the very fact that BiPS does show
a strong signal for the Wiener filtered map, mean that at some level
BiPS is sensitive to galactic residuals.

In summary, we study the Bipolar power spectrum (BiPS) of CMB which is
a promising measure of SI.  We find null measurements of the BiPS for
a selection of full sky CMB anisotropy maps based on the first year of
WMAP data. Our results rule out radical violation of statistical
isotropy in the CMB anisotropy measured by WMAP.

\end{document}